\documentclass{aa}
\usepackage{graphicx}
\graphicspath{{./}}

\usepackage{txfonts}
\newcommand{\be}{\begin{equation}}
\newcommand{\ee}{\end{equation}}
\newcommand{\etal}{et al.\ }
\newcommand{\msun}{\mbox{M$_{\sun}$}}
\newcommand{\msunpctwo}{\mbox{\, M$_{\sun} \, {\rm pc}^{-2}$}}
\newcommand{\natd}[2]{\mbox{$#1 \cdot 10^{#2}$}}
\newcommand{\pder}[2]{\frac{\partial #1}{\partial #2}}
\newcommand{\pdert}[1]{\pder{#1}{t}}

\begin{document}

\title{Are galactic disks dynamically influenced by dust?}
\author{Theis Ch.\inst{1,2} and Orlova N.\inst{1,3,4}}
  \institute{
  Institut f\"ur Theoretische Physik und Astrophysik d.\ Univ.\ Kiel,
    D--24098 Kiel, Germany\and
  Institut f\"ur Astronomie d.\ Univ.\ Wien, T\"urkenschanzstr.\ 17,
    A--1180 Wien, Austria,
    \email{theis@astro.univie.ac.at} \and
  Institute of Physics, Stachki 194, Rostov-on-Don, Russia,
    \email{orlova@rsusu1.rnd.runnet.ru} \and
  Isaak Newton Institute of Chile, Rostov-on-Don Branch}
\offprints{Ch.\ Theis (Vienna address)}
\date{received; accepted}

\abstract{Dynamically cold components are well known to destabilize
hotter, even much more massive components. E.g.\ stellar disks can become
unstable by a small admixture of cold gas or proto-planetary disks might
be destabilized by a small fraction of dust. In this paper we studied the 
dynamical influence of a cold dust component on the gaseous phase in the
central regions of galactic disks. We performed two-dimensional hydrodynamical
simulations for flat multi-component disks embedded in a combined 
static stellar and dark matter potential. The pressure-free dust component 
is coupled to the gas by a drag force depending on their velocity difference.

   It turned out that the most unstable regions are those with either a low 
or near to minimum Toomre parameter or with rigid rotation, i.e.\ the central 
area. In that regions the dust-free disks become most unstable for 
high azimuthal modes ($m \sim 8$), whereas in dusty disks all modes have 
a similar amplitude resulting in a patchy appearance. The structures in 
the dust have a larger contrast between arm and inter-arm regions than
those of the gas. The dust peaks are frequently correlated
with peaks of the gas distribution, but they do not necessarily coincide 
with them.
Therefore, a large scatter in the dust-to-gas ratios is expected.
The appearance of the dust is more cellular (i.e.\ sometimes connecting
different spiral features), whereas the gas is organized in a multi-armed
spiral structure.

   We found that an admixture of 2\% dust (relative to the mass of the gas)
destabilizes gaseous disks substantially, whereas dust-to-gas ratios below
1\% have no influence on the evolution of the gaseous disk. For a
high dust-to-gas ratio of 10\% the instabilities reach a saturation
level already after 30 Myr. The stability of the gaseous disks also strongly
depends on their Toomre parameter. But even in hot gaseous disks a
destabilizing influence of the dust component has been found.

}

\maketitle


\section{Introduction}
\label{introduction}

Recent high resolution
optical and sub-mm observations of the central regions of galaxies
have revealed that the circumnuclear regions contain
stellar-gaseous mini-disks with a size of about a few hundred pc.
An extensive observational program on HST by Carollo \etal
(\cite{carollo97}, \cite{carollo98a}, \cite{carollo98b})
who studied the nuclear regions of seventy five early and intermediate
type disk galaxies, unveiled an astonishing richness of structure in the 
nuclear regions of galaxies. About sixty percent of their sample have
mini-bars, spiral-like dust lanes, star-forming rings and spiral arms. 
Many nuclear regions are also well described as patchy or multi-armed.
Further investigations by Regan \& Mulchaey (\cite{regan99})
showed that spirals are the most common morphological structures in
the central regions of galaxies. Often the spiral patterns are not 
associated with the outer grand-design spiral patterns of these galaxies 
which points to different physical origins.

Revealing the nature of the mini-spirals in the
central regions of galaxies is of great importance for our understanding 
of a variety of astrophysical processes such as the mass accretion rate
of the central engines of galaxies or the evolution of the galactic disks
themselves.  Therefore, the dynamics of gaseous density waves in the central
region of disk galaxies has attracted more and more attention during the last
decade. Athanassoula (\cite{athanassoula92}) studied the gas flows in 
and around bars. She demonstrated by orbital analysis and
2D-hydrodynamical simulations that the dynamical properties of the 
disk in connection with a rotating large-scale bar, especially the existence 
of different periodic orbit families (or equivalently the existence of inner 
Lindblad resonances (ILR)) are key criteria for the existence and
shape of dust lanes. If $x_2$ and $x_3$-orbits are existing,
gas (and dust) might follow first $x_1$-orbits, until they enter regions
which support also ''anti-bar'' orbits. In these turbulent regions the gas 
dissipates energy and switches to $x_2$ or $x_3$-orbits which brings the
gas to the central area.

Different to the interpretation based on a stellar orbital analysis
Englmaier \& Shlosman (\cite{englmaier00}) suggested that 
the mini-spirals in the central regions of galactic disks
are related to the formation of grand-design spiral patterns in galaxies.
They argued that gas density waves -- different to stellar density waves --
are not completely damped or absorbed at the ILR and, thus, they may 
generate spiral structures at all radii including the nuclear regions. 
Such a model might explain the continuity of the spiral features at small
and large radii as well as the low arm-interarm contrast observed
in galactic centers. This scenario
is supported by their numerical simulations studying 
the gas response to a background gravitational potential composed of a
galactic disk and a large-scale stellar bar.

An alternative application of barred potentials as driving engines for 
the generation of nuclear spirals is to invoke secondary bars 
or small nuclear bars located well inside the ILR.
Wada \& Koda (\cite{wada01}) performed a set of multi-phase
hydrodynamical simulations studying the dynamics of
central mini-disks influenced by the potential of a weak mini-bar.
They took into account the self-gravity of the gas as well as cooling
and heating processes. They found the formation of filaments and cusps
on a parsec-scale which resemble the observed morphological patterns in 
the central regions of galaxies. 
Another good example for the influence of a nuclear bar is the match
between observations and numerical (SPH-)models by Ann (\cite{ann01}). 
He reproduced well the shape and orientation of the nuclear ring of the 
barred galaxy NGC 4314.
Jogee \etal (\cite{jogee02}) studied the propagation of density waves 
triggered by bar shocks which were derived from multiwavelength observations 
of NGC 5248. Motivated by the existence of massive molecular arms inside 
the disk they incorporated the self-gravity of the gas. Their comparative 
analysis shows an agreement between the modelled and observed gas morphology, 
gas kinematics, and pitch angle of the spirals. These results argue for a 
dynamical coupling between the nuclear region and the surrounding disk. 

A common property of all the mentioned mechanisms is that they usually 
result in structures dominated by two or a few arms. 
E.g.\ a nuclear bar would naturally induce a
two-armed structure. Similarly, grand-design features penetrating 
from the outer disk through the ILR will also keep their symmetry
properties characterized by low azimuthal mode numbers.
However, Elmegreen \etal (\cite{elmegreen02}) found that nuclear dust spirals
differ from main-disk spirals in several respects:
the nuclear spirals are not associated with star formation, they are 
very irregular with both trailing and leading components that often cross,
and they completely fill the inner disk with a constant surface density.

Little is yet known about the dynamics on the 10-100 pc scale 
of the central regions of galaxies.
Attempts to explain the existence of such structures as a result of 
gravitational instability of the dynamically decoupled mini-disks were 
believed to be unsuccessful because of the low surface densities of the fast 
spinning gaseous mini-disks. 
As an alternative to gravitational instability as an origin of structure
in mini-disks 
Elmegreen \etal (\cite{elmegreen98}) suggested that mini-spirals may be
the result of an acoustic amplification of the grand-design
spirals propagating to the central regions of galaxies. Montenegro \etal 
(\cite{montenegro99}) demonstrated that such an amplification mechanism 
can indeed work inside the ILR or outside the OLR. 

All discussed mechanisms have some weak points and can not fully 
account for all the observational data. We consider a new approach 
to understand the origin of mini-spirals in the central regions of galaxies.
Observations show that the circumnuclear disks of galaxies are dusty,
and dust may play an important r\^ole in their dynamics. It is known that 
cold components greatly destabilize multi-component gravitating
disks (Jog \& Solomon \cite{jog84}, Bertin \& Romeo \cite{bertin88},
Orlova \etal \cite{orlova02}). An admixture of a dynamically cold dust 
component to the gaseous mini-disks of galaxies might therefore strongly 
destabilize the disk leading to the formation of spiral structure.  
From a stability analysis Noh \etal (\cite{noh91}) showed
that a dust component can strongly destabilize proto-planetary disks. 
The admixture of only 2\% of dust can enhance the growth rates
of the dominant gaseous phase significantly. This effect becomes
even larger with less massive disks. A conservative estimate of the
dust-to-gas ratio in the solar neighborhood gives an average value of 0.6\% 
ranging from 0.2\% up to 4\% in $H_2$ regions (Spitzer \cite{spitzer78}). 
These are lower limits because large grains which contain much mass do 
not have any detectable extinction. Recent observational data 
by Maiolino \etal (\cite{maiolino01}) exhibit evidence for anomalous 
properties of the dust grains in the galactic nuclei. By comparing the 
reddening of optical and infrared broad lines and the X-ray absorbing 
column density they found that the $A_V/N_H$ ratio is
significantly lower in the circumnuclear regions of galaxies
than in the diffuse regions of galaxies. This indicates that
the dust composition in the circumnuclear regions of galaxies
could be dominated by large grains and, thus, substantial amounts
of dust might have been undetected so far.

In this paper we study whether the origin of mini-spirals
can be explained by a destabilizing r\^ole of a dust component
in the circumnuclear disks. We restrict our analysis here 
to systems without a nuclear or a large scale bar. Also we do not
consider here any formation or destruction processes of dust. Thus,
we investigate the impact of a frictional force exerted by the
interaction of dust and gas on the dynamics of galactic nuclear regions.
This investigation is done by means of 2-dimensional multi-component 
hydrodynamical simulations. In the next Section we describe the numerical
model, i.e.\ the basic hydrodynamical equations and the dust-gas
interaction as well as our numerical code. In Sect.\ \ref{results}
the results of the numerical simulations are shown which are discussed in
more detail in Sect.\ \ref{discussion}. Finally, a summary is given in 
Sect.\ \ref{summary}.


\section{Numerical method}
\label{numericalmethod}
%
%
\subsection{Pure hydrodynamics}
\label{purehydrodynamics}

  We studied numerically the hydrodynamical equations for a 2-dimensional
single- or multi-component disk. Thus, we solved the continuity equation
\begin{equation}
    \pdert{\Sigma_{g,d}} +
         \nabla \cdot (\Sigma_{g,d} \vec{v}_{g,d}) = 0
\end{equation}
and the momentum equations for gas and dust. The momentum equations read
for gas
\begin{equation}
    \pdert{\vec{v}_g} + (\vec{v}_g \cdot \nabla) \vec{v}_g
        + \frac{\nabla P_g}{\Sigma_g}
        + \nabla (\Phi + \Phi_{\rm HBSD}) = S_g(\vec{v}_g) 
\end{equation}
and for dust
\begin{equation}
    \pdert{\vec{v}_d} + (\vec{v}_d \cdot \nabla) \vec{v}_d
        + \nabla (\Phi + \Phi_{\rm HBSD}) = S_d(\vec{v}_d) \,\, .
\end{equation}
The components are
denoted by $g$ for the gaseous phase and by $d$ for the dust.
$\Sigma_{g,d}$ are the surface densities and $\vec{v}_{g,d}$ the
velocities. $P_g$ is the pressure of the gas which is given in the
2d-case as force per unit length.
$\Phi$ denotes the potential of the self-gravitating disk. 
It is derived from the Poisson equation 
\begin{equation}
  \Delta \Phi = 4 \pi G \Sigma(R,\varphi) \,\, \delta(z) 
              = 4 \pi G \left( \Sigma_g + \Sigma_d \right) \delta(z)\,\, .
\end{equation}
An external potential is added by a stationary contribution 
$\Phi_{\rm HBSD}$ related to the halo, the bulge and/or a stellar
disk component. These external potentials are chosen to match -- together
with the potential of the disk -- a given rotation curve. 

The main difference between the gaseous and the dust component is that
the dust is treated as a pressureless phase. Therefore, if
gas and dust are in rotational equilibrium, there is a velocity
difference between both components which might give rise to a 
non-negligible frictional force depending on the cross-section for the gas-dust
interaction. This interaction is described by the source terms $S(\dots)$
on the RHS of the hydrodynamical equations. Since we do not
consider dust formation and destruction processes, the source terms
in the continuity equation vanish. However, frictional terms show up
in the equations of motion. The dust implementation will be described in the
next paragraph in detail.

  The set of hydrodynamical equations is closed by a polytropic
equation of state
\begin{equation}
    P_g = K \Sigma_g^{\gamma_g} \,\,\, .
    \label{eqstate}
\end{equation}
For the gaseous phase we set the polytropic exponent to $\gamma_g=5/3$.
The constant $K$ is chosen to yield a given minimum Toomre parameter.

%
%
\subsection{Treatment of the dust component}
\label{dusttreatment}

Gravitating mini-disks in galaxies
differ from ''normal'' galactic disks in which the cold component (gas)
is coupled gravitationally to the dynamically hot component (stars).
The main difference between galactic disks and
mini-spirals is that the cold component (dust) in the mini-disks
is not only coupled by gravity to the hotter component (gas), but 
also by a frictional force between both components.

  In order to include this drag, we added a source term to the
equations of motion following the general form suggested by Noh \etal 
(\cite{noh91})
\begin{eqnarray}
   \vec{f} \equiv S_d(\vec{v}_d) & = & - A (\vec{v}_d - \vec{v}_g) \\
   S_g(\vec{v}_g) & = & - \frac{\Sigma_d}{\Sigma_g} \, \vec{f} \,\,.
   \label{eqsourcegas}
\end{eqnarray}
The second source term, Eq.\ (\ref{eqsourcegas}) follows from the 
requirement of momentum conservation. The physics of the friction is 
enclosed in the frictional timescale $A^{-1}$. We implemented three 
different prescriptions of the frictional term in our models.

   The simplest approach is to assume a time- and 
position-independent frictional timescale $\tau_d$, i.e.\
\begin{equation}
      A = \tau_d^{-1} \,\,\, .
      \label{eqatimescale}
\end{equation}
Though this is a very rough assumption, it allows for a direct
comparison between the drag term and other dynamical quantities,
e.g.\ the dynamical timescale given by the rotation period. More
detailed approaches are based on a microscopic view of the frictional
process, i.e.\ the momentum exchange between gas and dust particles
by collisions and the equipartition of momentum within the gaseous phase. 
Two limiting situations are described in the following sections in 
more detail.

\subsubsection{Collisional time scale (equal disk heights)}
\label{collisionaltimescale}

Because the size \mbox{$R_d \sim 0.1 \mu{\rm m}$} 
of typical dust particles is much smaller than the mean molecular 
free path \mbox{$\lambda \sim 0.1 {\rm AU}$} of the ambient gas particles
in galactic central regions, the velocities of gas molecules
are uncorrelated with the velocity of the dust particle itself 
and, thus, the friction can be described
as an ``Epstein'' drag (Goodman \& Pindor \cite{goodman00}). 
The latter is derived from kinetic gas theory (Epstein \cite{epstein24}). 
Only for very large grains residing in very dense regions the Stokes' formula 
for the friction has to be used (then the friction is given by a laminar 
viscous flow over the dust particle).
Assuming ''Epstein'' friction is acting Noh \etal (\cite{noh91})
discerned between the two limiting cases of different scale heights of 
the dust component: either similar disk heights of gaseous and dust disk or
a much thinner dust disk. 

   In case of similar or equal disk heights
it is a reasonable ansatz to use the time scale $\tau_c$ for energy and
momentum exchange due to collisions between gas and dust particles as an 
estimate of $A^{-1}$
\begin{equation}
   A = \tau_c^{-1} = \frac{\sigma_c \rho_g v_{\rm th}}{m_d} \,\,\, .
   \label{eqacollision}
\end{equation}
$m_d$ denotes the mass of a dust particle, $\rho_g$ the mass density of the 
gas and $v_{\rm th}$ the thermal velocity of the gas. The collisional 
cross section $\sigma_c$ can be estimated from the geometrical cross section,
i.e.\ from the average radius $R_d$ of a dust particle
\begin{equation}
  \sigma_c = \pi \cdot R_d^2 \,\,\, .
\end{equation}
The collisional time scale is then given by
\begin{eqnarray}
   \tau_c & \approx & \natd{1.49}{3} \cdot
    \left( \frac{m_d}{10^{-14} \, {\rm g}} \right)
    \left( \frac{R_d}{0.1 \, \mu{\rm m}} \right)^{-2} \cdot \nonumber \\
          & & 
    \left( \frac{\Sigma_g}{10^3 \, \msunpctwo} \right)^{-1}
    \left( \frac{H}{100 \, {\rm pc}} \right) \cdot
    \left( \frac{v_{\rm th}}{10 \, {\rm km s}^{-1}} \right)^{-1}
     \,\, {\rm yrs}
    \label{eqacollisionnumbers}
\end{eqnarray}
$H$ is the full scale height of the gas, its spatial density is
estimated by $\rho_g = \Sigma_g / H$.

\subsubsection{Dynamical timescale (thin dust disk)}

  The other limiting case corresponds to a dust disk which is substantially
thinner than the gaseous disk. In that
case the time scale should be of the order of the dynamical
time scale given by the inverse circular frequency $\Omega^{-1}$
(Noh et al.\ \cite{noh91}), i.e.\
\begin{equation}
    A = \Omega \,\,\, .
   \label{eqadynamics}
\end{equation}
The basic assumption is here that gas and dust establish very fast 
collisional equilibrium in the thin dust layer and that the gas momentum 
is then mixed vertically in a sound travelling time scale.
When taking typical values we see from Eq.\ (\ref{eqacollisionnumbers}) 
that the collisional time scale $\tau_c$ is indeed much shorter than the 
dynamical timescale. E.g.\ the rotation period in the inner regions of 
galaxies is about $10^7$ yrs.
The sound propagation time $\tau_s$ can be estimated from
$\tau_s \sim H_g / v_{\rm th}$ ($H_g$ is here the disk scale height of
the gaseous phase). In dynamical equilibrium the scale height is
related to the velocity dispersion by $H_g \sim v^2_{\rm th} / (\pi G \Sigma)$ 
(see e.g.\ Binney \& Tremaine \cite{binney87}, hereafter BT87) which yields
\begin{eqnarray}
    \tau_s \sim \frac{v_{\rm th}}{\pi G \Sigma} 
           \sim \kappa^{-1} \sim \Omega^{-1}
    \nonumber
      \,\,\, . 
\end{eqnarray}
The second relation is derived from the Toomre stability criterion. The last
relation between the circular frequency $\Omega$ and the epicyclic frequency
$\kappa$ holds for all reasonable rotation curves within a factor of 2.
Hence, in the limit of a thin dust disk the friction time scale is 
dominated by the sound propagation time which is of the order of the
dynamical time.


\subsection{Numerical implementation}
\label{numericalimplementation}

The nonlinear analysis implies the solution 
of the full set of hydrodynamical equations. For this purpose, we developed
a two-dimensional numerical code which is similar to the ZEUS-2D code 
by Stone \& Norman (\cite{stone92}). The hydrodynamical equations are 
discretized on a Eulerian grid in polar coordinates. The different
terms are treated by operator splitting. Advection is performed by a
second order Van Leer advection scheme. 

In radial direction we use a logarithmic grid which allows a
very high resolution in the central region. Therefore, a radial resolution 
of 128 cells should be sufficient for two-armed spirals
as Englmaier \& Shlosman (\cite{englmaier00}) demonstrated. 
In azimuthal direction, however, 128 grid cells can be 
critical, because the maximum growth rate might be found for high 
values $m$ of the azimuthal wavenumber. In case of protoplanetary
disks Noh \etal (\cite{noh91}) found $m\sim 5\dots10$ to be the most 
unstable modes. In order to accommodate such high modes, we used mainly 
a grid size of 270$\times$270 cells\footnote{For our simulations on the 
NEC-SX5 of the Kiel computing center we used a 270$\times$270 grid which has a 
similar physical resolution as the ''normal'' 256$\times$256 grid, 
but a much better computational performance of the 
vector-optimized FFT routine supplied with NEC's MATHKAISAN library. 
The increase of performance is caused by avoiding bank conflicts in the 
memory access on the NEC-SX5.}. The physical sizes of the cells
are similar in azimuthal and radial directions at all distances. 
The typical radial extent of the whole grid covers radial ranges from a 
few 10 pc to a few kpc. The Poisson equation is solved by 
applying the two-dimensional Fourier convolution theorem in polar 
coordinates (BT87).
The hydrodynamical timestep is calculated by a combination of the standard
Courant-Friedrichs-Levy criterion and the timescale is 
derived from the frictional force. 


\subsection{Units}
\label{units}

   For our simulations we set the gravitational
constant to unity and we chose the length unit to 1 kpc and the
mass unit to $10^9 \msun$. The system time is then measured in
\natd{1.49}{7} yrs, the velocity in 65.6 \mbox{km\,s$^{-1}$}, the
circular speed in 65.6 \mbox{km\,s$^{-1}$ kpc$^{-1}$} and
the surface density in $10^3 \msunpctwo$. If not stated explicitly,
all quantities are given in these units.


\subsection{Initial models}
\label{initialmodels}

    For most of our simulations the mass distribution of the disk is 
described by a weakly perturbed exponential profile. The perturbation
is created by multiplying the surface density with a factor
$(1+C r)$. $r$ is a random number in the interval \mbox{[-1,1]} 
and the amplitude $C$ is typically selected to yield a global 
Fourier amplitude of the order $10^{-6}$ (cf.\ also Eq.\ (\ref{eqamplitude})). 

The initial velocities are derived from rotational equilibrium, i.e.\ there is
no initial radial motion. The rotation curve is specified by 
\begin{equation}
   v_c(R) = v_\infty \cdot \frac{\displaystyle \frac{R}{R_{\rm flat}}}
    {\left[ 1 + \left(\displaystyle 
                 \frac{R}{R_{\rm flat}} \right)^{n_t} \right]^{1/n_t}}  \,\, .
   \label{eqvcirc}
\end{equation}
The radius $R_{\rm flat}$ defines the transition between the central rigid
rotation and an outer flat rotation curve. The sharpness of the transition
is controlled by $n_t$. The azimuthal velocity of the gaseous phase
is calculated by the (frictionless) Jeans' equation for the radial velocity 
component which reads in cylindrical coordinates
\begin{equation}
   \pdert{u_g} + u \pder{u_g}{R} + \frac{v_g}{R} \pder{u_g}{\phi} 
              - \frac{v_g^2}{R} =
        - \frac{1}{\Sigma_g} \cdot \pder{P_g}{R}
        - \pder{}{R}\left( \Phi + \Phi_{\rm HBSD}\right) \,\, .
   \label{eqjeansu}
\end{equation}
\noindent
In case of the initial equilibrium, the radial velocity $u_g$ vanishes
and, hence, the azimuthal velocity $v_g$ is given by
\begin{equation}
   v_g^2 = \frac{R}{\Sigma_g} \cdot \pder{P_g}{R} +
           R \pder{}{R}\left( \Phi + \Phi_{\rm HBSD}\right)     
         = v_c^2(R) - v_P^2(R) \,\, .
  \label{eqvazimuthal}
\end{equation}
\noindent
where we used that the rotation curve is derived from the gravitational 
potential, i.e.\ $v_c^2(R) = R \pder{}{R}(\Phi + \Phi_{\rm HBSD})$,
and the pressure contribution $v_P$ to the rotation curve is abbreviated as
\begin{equation}
   v_P^2 \equiv - \frac{R}{\Sigma_g} \cdot \pder{P_g}{R} \,\, .
   \label{eqvpressure}
\end{equation}
\noindent
In case of the pressureless dust component its initial azimuthal speed 
$v_d$ is directly given by the rotation curve (\ref{eqvcirc}).

From the mass and potential distribution (or the rotation curve), 
the radial shape of the profile of the Toomre parameter
\begin{equation}
   Q \equiv \frac{v_{\rm th} \kappa}{\pi G \Sigma_g}
\end{equation}
can be derived except for a constant factor. This factor can be specified 
by requiring a minimum value $Q_{\rm min}$ for the Toomre parameter.


\subsection{Fourier Modes}

In order to quantify the (global) stability of the disk we use 
the global Fourier amplitudes of each component
(cf.\ e.g.\ Laughlin et al.\ \cite{laughlin98})
\begin{equation}
     C_m  \equiv {1 \over {M_{\rm disk}}}
     \left| \displaystyle \int_{0}^{2 \pi} \int_{R_{\rm in}}^{R_{\rm out}}
     \Sigma(r,\phi) r dr \, e^{-im\phi} d\phi \right| \,\,\,\,\,
     \mbox{\rm ($m>0$).}
     \label{eqamplitude}
\end{equation}
\noindent
$M_{\rm disk}$ is the mass of the disk for the component of interest
in the specified radial interval $[R_{\rm in},R_{\rm out}]$.
$\Sigma$ denotes the corresponding surface density. Global Fourier
amplitudes are calculated by integrating over the whole disk.

  Additionally, a mode $m=0$ is introduced which quantifies the
radial mass redistribution with respect to the initial one:
\begin{equation}
   C_0 \equiv \frac{2\pi}{M_{\rm disk}} \int_{R_{\rm low}}^{R_{\rm hig}}
         |\bar{\Sigma}(R,t) - \bar{\Sigma}_0(R)| R dR \,\, .
\end{equation}
$\bar{\Sigma}(R,t)$ and $\bar{\Sigma}_0(R)$ are the azimuthally averaged
surface densities at time $t$ and at the begin of the simulations, 
respectively.
As long as the evolution is well described by linear perturbation theory, 
no substantial radial mass transport occurs. Thus, the increase of $C_0$ is
a good indicator for the onset of non-linear effects during the growth of
perturbations.

\subsection{A single component model}
\label{singlecomponentmodel}

\begin{figure}
   \resizebox{\hsize}{!}{
     \includegraphics[angle=90]{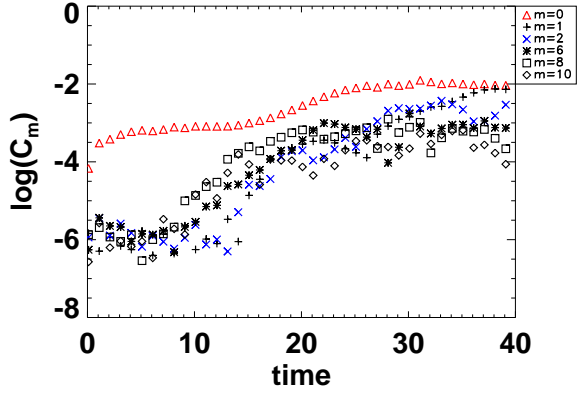}
   }
   \caption{Temporal evolution of the Fourier amplitudes of 
    the $m=0,1,2,6,8,10$-modes for a single component model with the
    mass distribution of the gaseous component in the reference model.
    The time unit is $\natd{1.5}{7}$ yrs.}
   \label{singlecomponent_globalmodes}
\end{figure}

\begin{figure*}[tp]
  \centerline{\hbox{
  \includegraphics[angle=90,width=8.5cm]{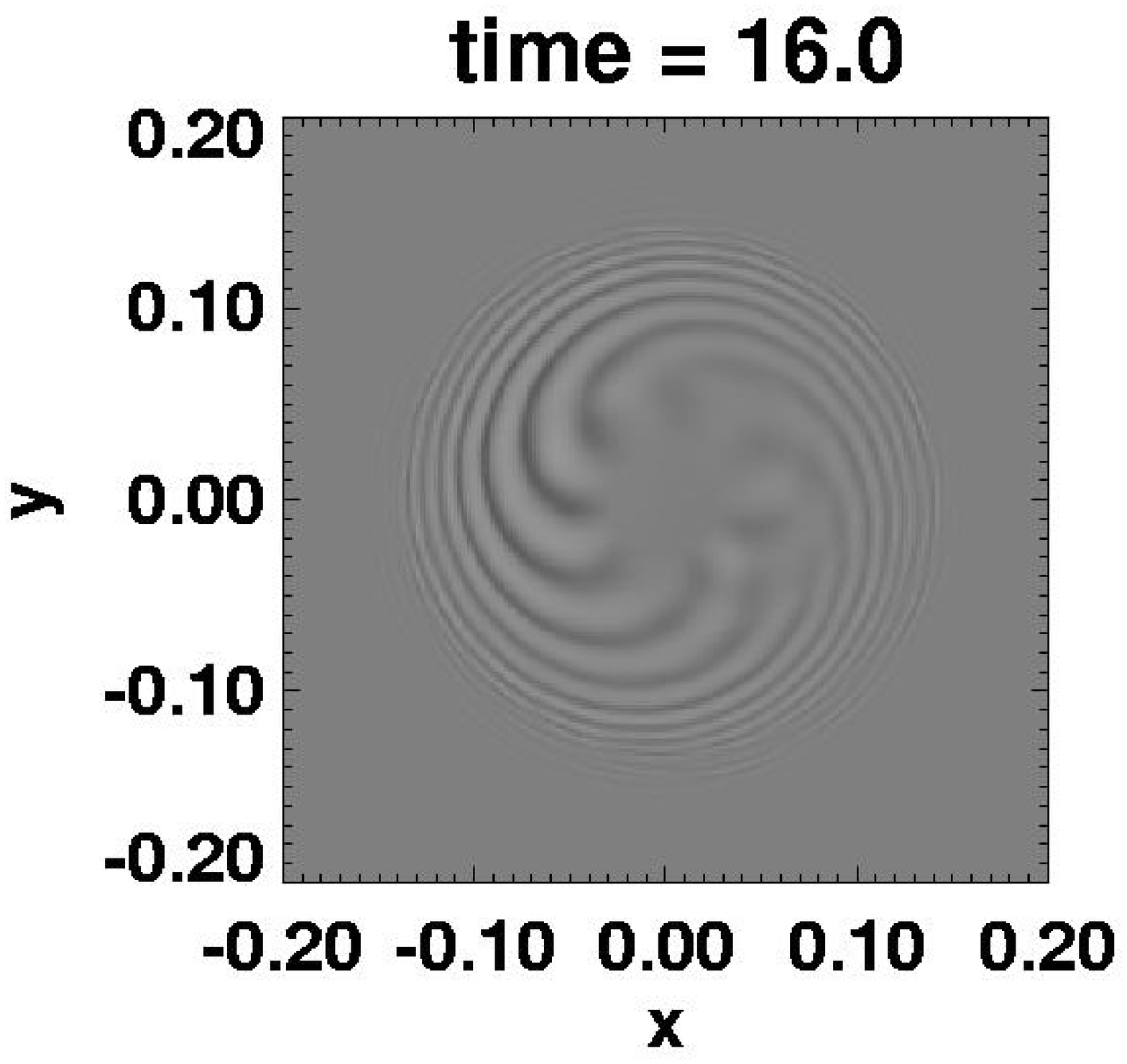}
  \includegraphics[angle=90,width=8.5cm]{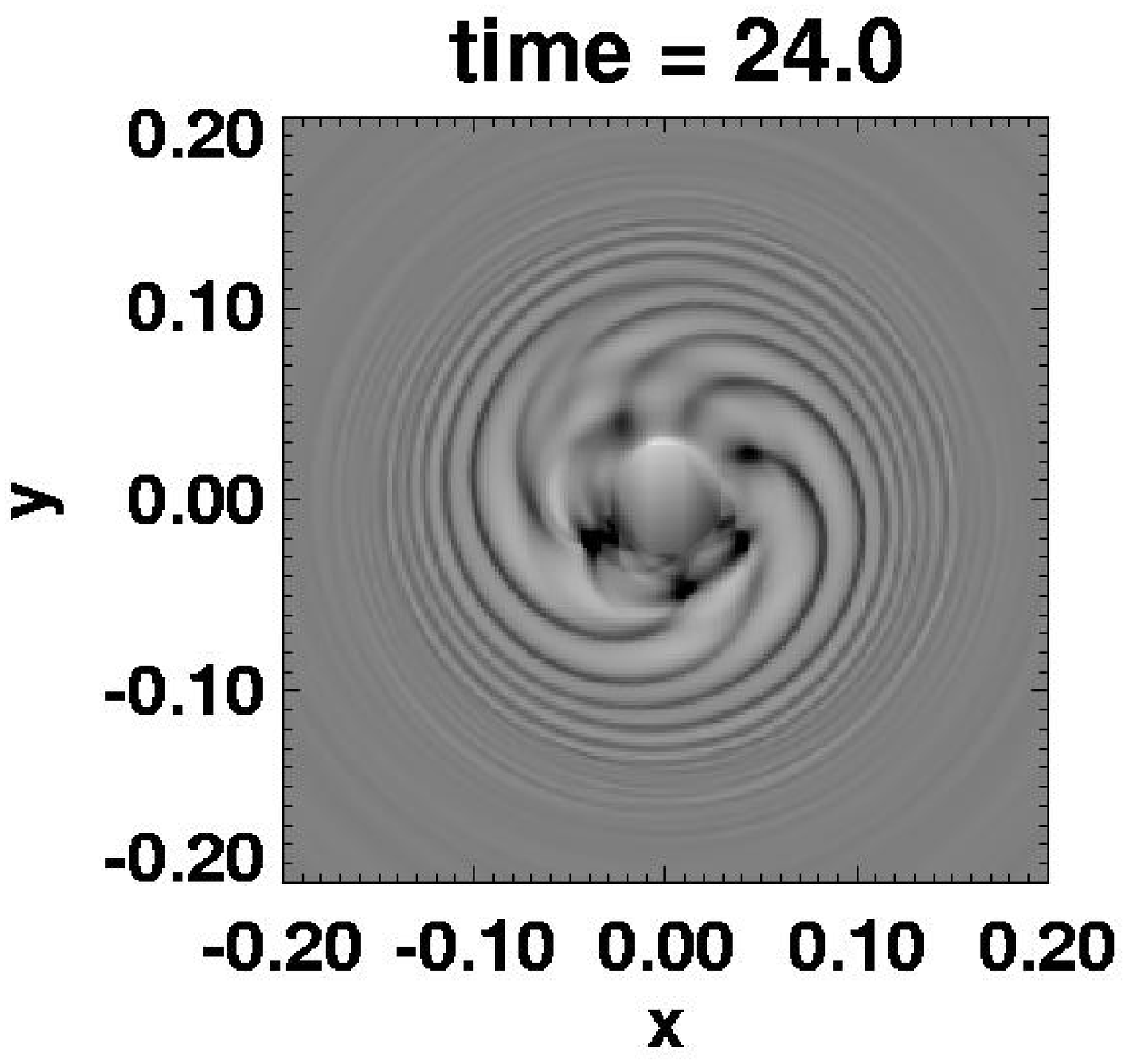}
  }}
  \vspace*{0.3cm}

  \centerline{\hbox{ 
  \includegraphics[angle=90,width=8.5cm]{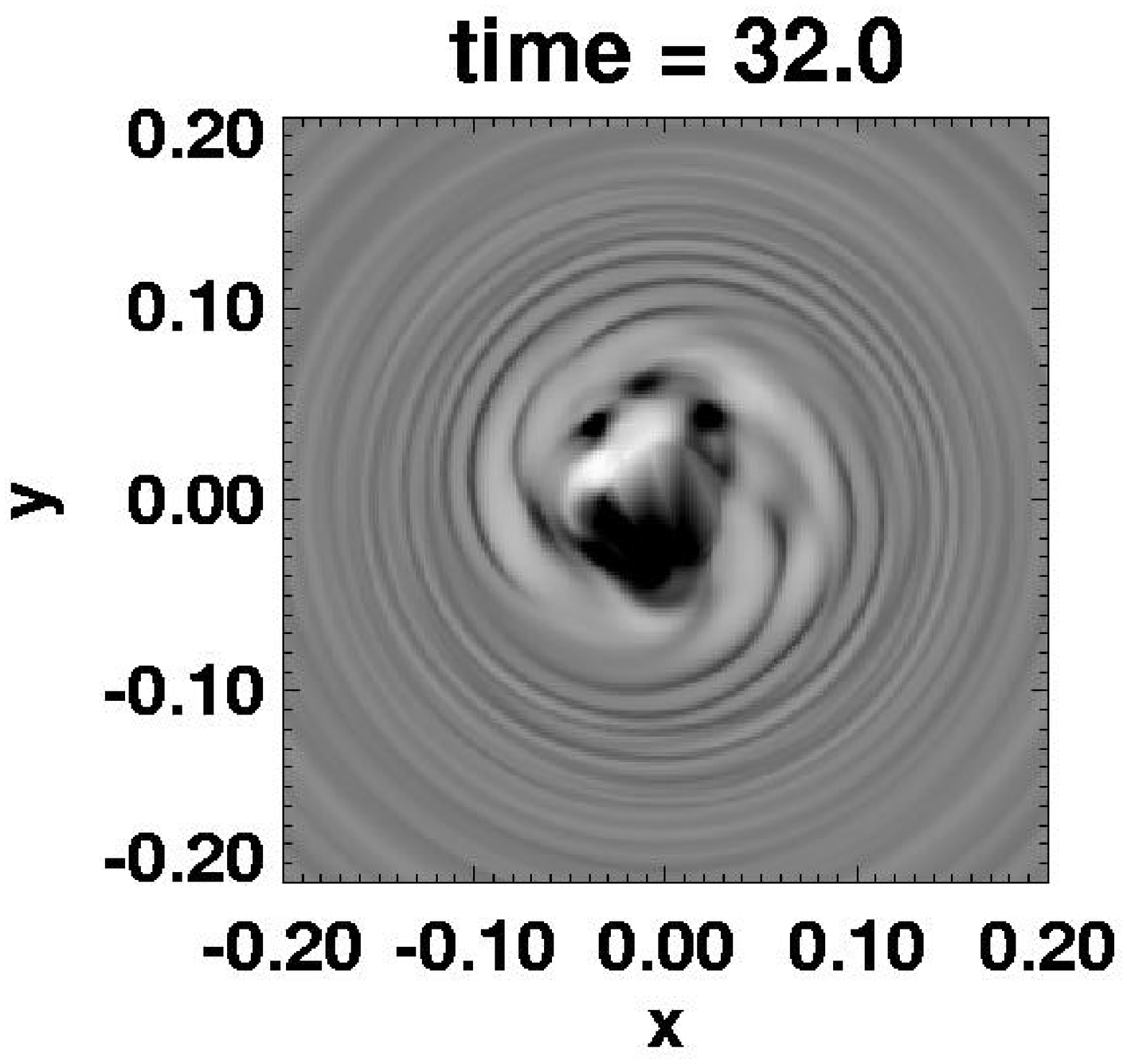}
  \includegraphics[angle=90,width=8.5cm]{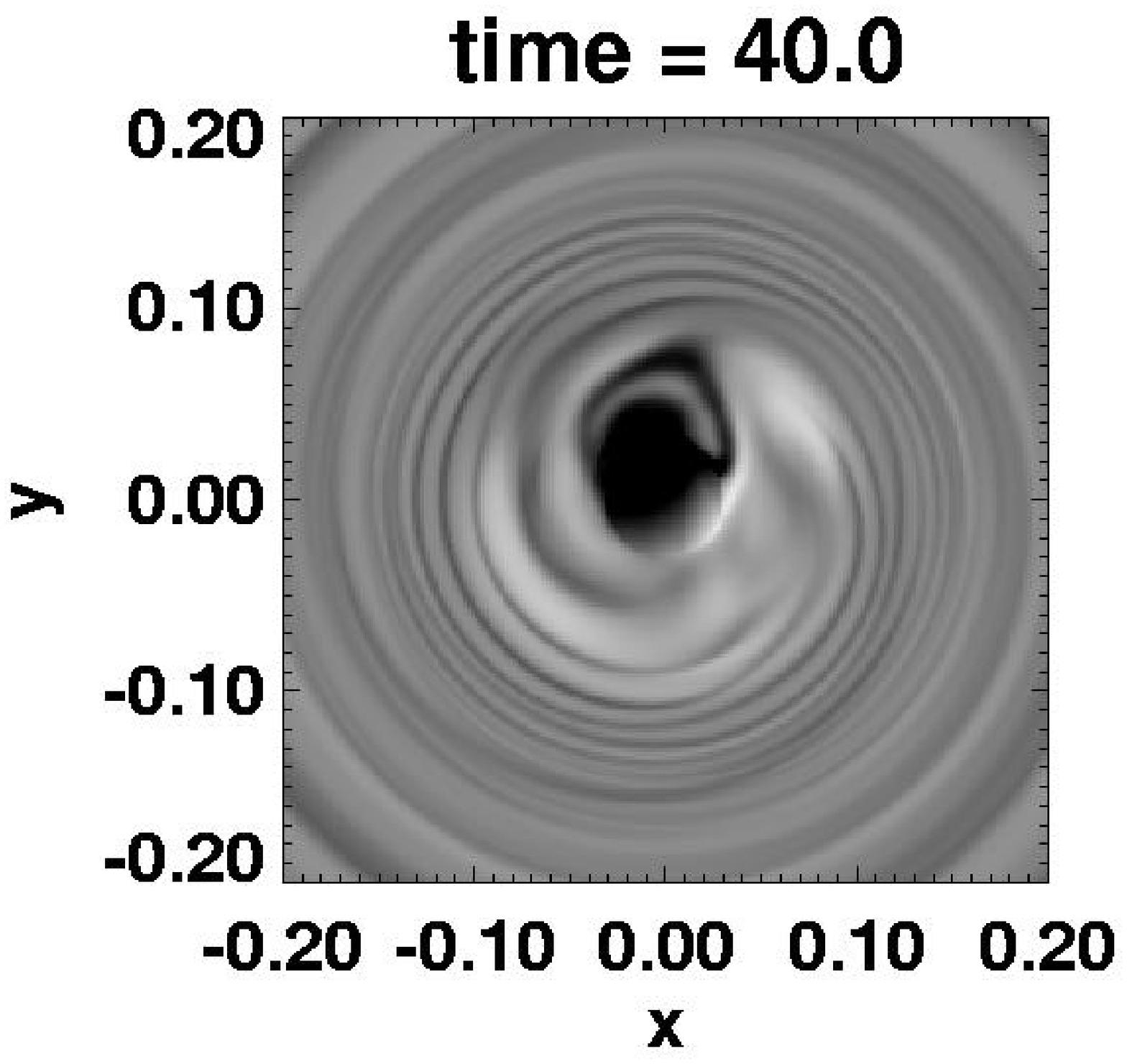}
  }}
  \caption{Spatial distribution of the surface density perturbations
    (normalized to the their initial values) of the single component
     gaseous nuclear disk ($Q_{\rm min} = 1.54$) at different times:
     during the linear growth regime ($t=16$, upper left), 
     saturation of the higher-order moments ($t=24$, upper right), 
     saturation of the lower-order moments ($t=32$, lower left) 
     and at the end of the simulation ($t=40 \sim 600$ Myr, lower right).
    Areas devoid of material are white, areas with a density enhancement 
    of a factor of 2 or more are black. The grey area seen at the outer 
    edges correspond to no deviation from the initial surface density.
    Note that the data inside the inner boundary of the computational grid
    ($|x|,|y| \leq 0.03 = 30$ pc) are artificial. 
    They are created by the graphics software when 
    converting from the polar grid to the Cartesian grid of the image.
   \label{singlecomponent_denspert_image}
   }
\end{figure*}

\begin{figure}[tp]
  \resizebox{\hsize}{!}{
   \includegraphics[angle=90,width=8.5cm]{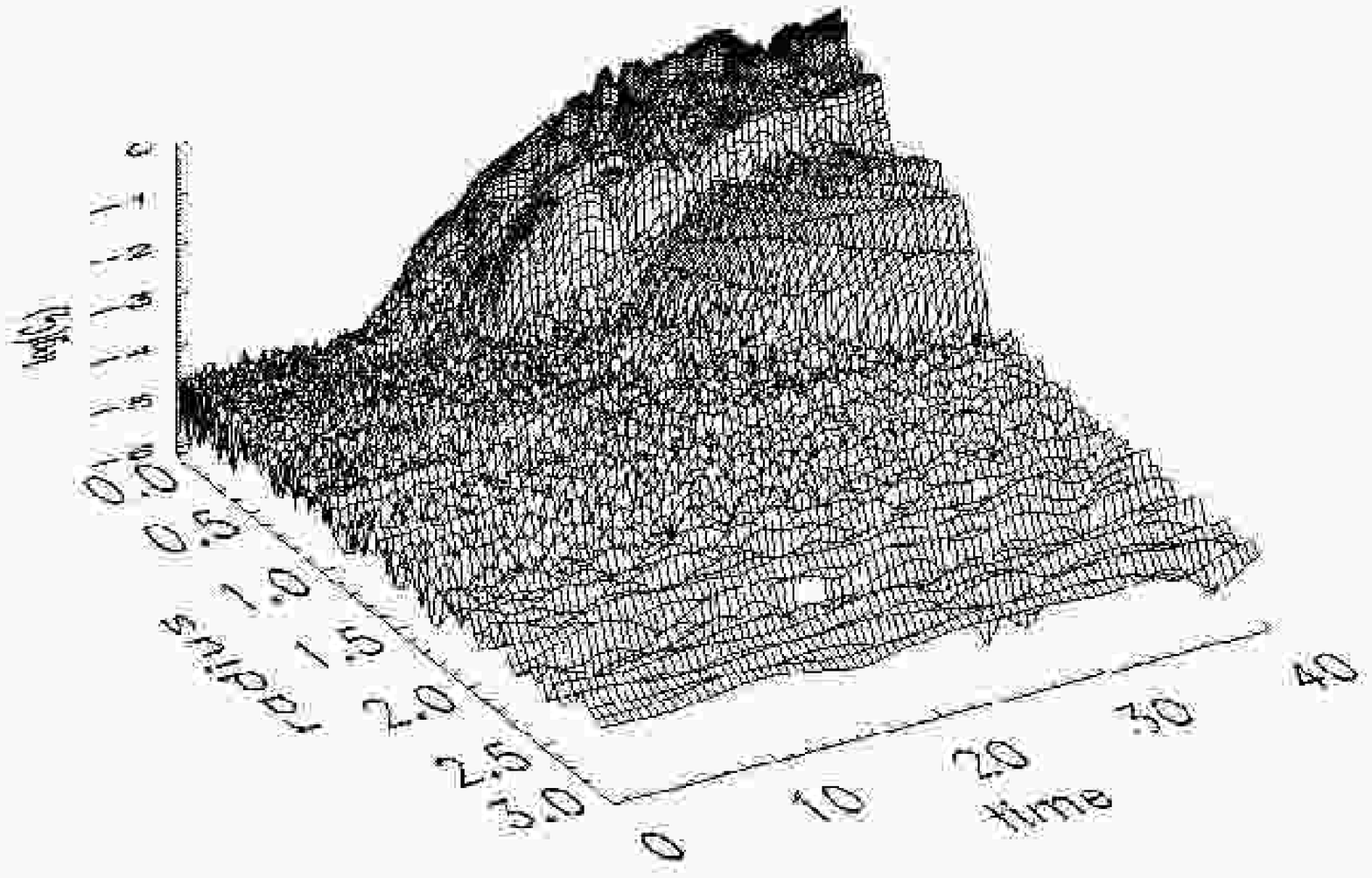}
   }
  \resizebox{\hsize}{!}{
   \includegraphics[angle=90,width=8.5cm]{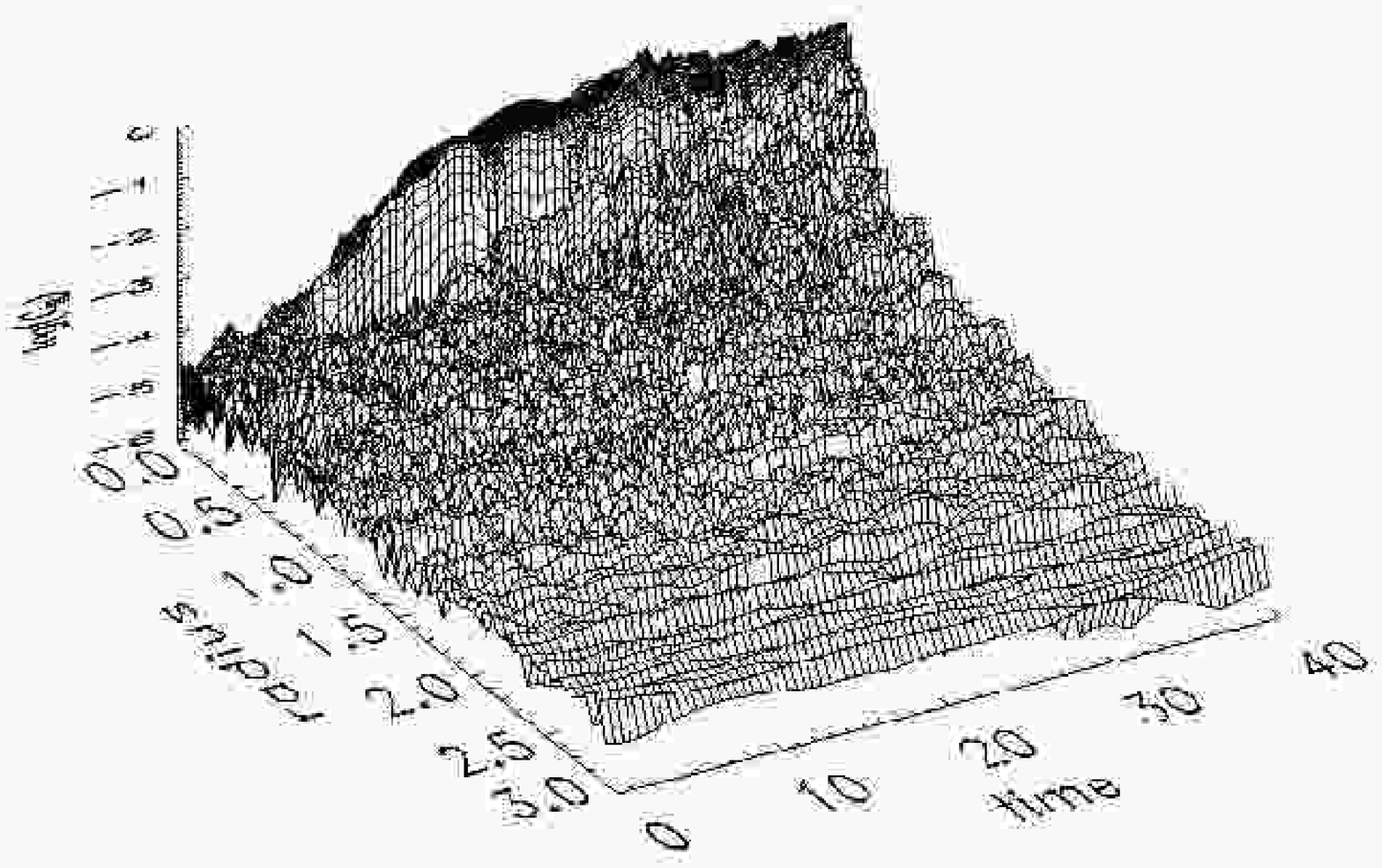}
   }
  \caption{Temporal evolution of the spatially resolved Fourier
    amplitudes for the $m$=2-mode (upper diagram) and the
    $m$=8-mode (lower diagram) for a single component model with the
    mass distribution of the gaseous component in the reference model.
    The amplitudes are calculated according to Eq.\ (\ref{eqamplitude}) 
    integrating over small radial annuli defined by the numerical
    grid resolution. The time unit is $\natd{1.5}{7}$ yrs.}
   \label{singlecomponent_spat_ampl}
\end{figure}

   In order to study the properties of a standard disk, but also to
test the code for our application we performed a series of single-component 
(i.e.\ purely gaseous) simulations based on the mass distribution of a 
reference model. Its mass distribution is characterized by a submaximal 
exponential disk and a rotation curve which increases linearly in the 
center and is flat outside (for exact values see the next section).

In order to quantify the structural evolution we first discuss
the global Fourier amplitudes (Fig.\ \ref{singlecomponent_globalmodes}):
they remain constant on their initial level of about $10^{-6}$ during the 
first 100 Myrs (or $t\approx 7$). Then the high-order
modes $m=6,8,10$ start to grow exponentially reaching a saturation level
of about $10^{-3}$. The fastest growing and dominant mode is the 
\mbox{$m$=8}-mode. This is also seen in the image of the density perturbations
(Fig.\ \ref{singlecomponent_denspert_image}, upper row). The wavelengths
of the perturbations are rather short as expected from the small critical 
wavelength $\lambda_c \equiv 4\pi^2 G \Sigma / \kappa^2 \approx 100$ pc.
From linear theory the maximum growth is expected for a wavelength
at about $\lambda_c / 2 \sim 50$ pc (see BT 87). Later than 
the high order modes, the amplitudes of the low-order modes, 
especially $m=2$ and $m=1$, commence to increase with a time delay of 100 Myr.
At the end of the simulation at $t=40$ (600 Myr) their saturation levels 
slightly exceed the saturation levels of the high order modes.
The overall appearance is then dominated by a mainly lopsided and
very irregular morphology (Fig.\ \ref{singlecomponent_denspert_image}, 
lower row).  

The radial mass
distribution varies during the first 600 Myr only within the innermost
100 pc which is reflected also in the evolution of the $m$=0-mode:
After a fast readjustment of the radial mass profile according to the
initial density perturbations, it remains constant
until $t\sim 15 \approx 225$ Myr. 
Then a radial mass transfer sets in, which stops at about
$t\sim 30 \approx 450$ Myr. Though the values of the perturbation
amplitudes are rather small, this does not mean that the response of the
disk is everywhere in the linear regime. The outer regions remain
almost unevolved, whereas the inner regions undergo strong 
morphological changes. It is only the ''standard'' normalization to the 
whole disk mass which keeps the global Fourier amplitudes rather small.
In order to demonstrate this, Fig.\ \ref{singlecomponent_spat_ampl}
shows the temporal evolution of the spatially resolved Fourier
amplitudes for $m$=2 and $m$=8. As seen already for the global modes,
the high-order modes start to grow first. Additionally, the growth
begins in the central area where the surface densities are at maximum
and the dynamical timescales are short. Throughout the whole simulation
the region of growing high-order modes is restricted to the central 250 pc.
Only the $m$=2-mode expands after $t\sim 20 \approx 300$ Myr to 
distances up to 1 kpc, 
but still with amplitudes much smaller than in the center. In the 
nuclear region both, low- and high-order modes, reach a non-linear 
saturation level of about 5-20\%.

  From a technical point of view, the conservation of various quantities
is quite well fulfilled. At the end of the simulation (after 
$\sim \natd{1.8}{5}$ numerical steps) the mass is conserved within 
machine accuracy of double precision numbers ($\sim 10^{-16}$). 
Energy is conserved to better than $\natd{5}{-5}$ and angular momentum 
better than $\natd{2}{-6}$ (the numbers denote the relative deviations 
from their initial value). This simulation was performed without 
artificial viscosity.


\section{Results}
\label{results}

   In this section we present the results for dusty disks. First,
we discuss the properties of a reference model in Sect.\ 
\ref{referencemodel}. Afterwards the influence of various parameters
is shown in Sect.\ \ref{parameterstudy}.


\subsection{The reference model}
\label{referencemodel}

\subsubsection{The start model}  

\begin{figure}
   \resizebox{\hsize}{!}{
     \includegraphics[angle=270]{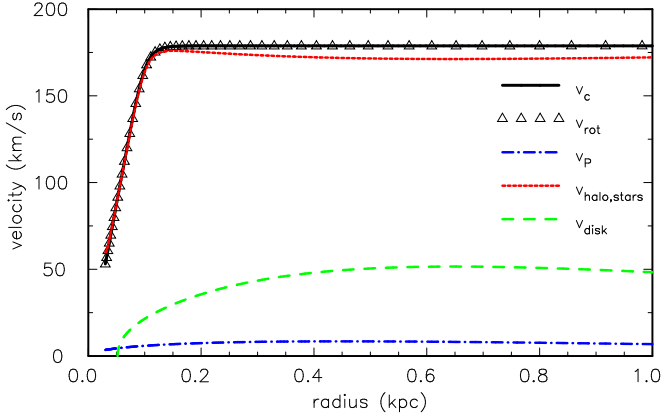}
   }
   \caption{Radial profile of the initial rotation curve and the
    different contributions to the gaseous azimuthal velocity of
    the reference model: rotation curve, i.e.\ circular velocity
    $v_c$ (solid), azimuthal velocity of gas $v_{\rm rot}$ (triangles),
    pressure contribution to rotational equilibrium $v_P$ (dot-dashed),
    halo (and stellar) contribution to rotational equilibrium
    $v_{\rm halo,stars}$ (short-dashed) and contribution form 
    the self-gravity of the disk $v_{\rm disk}$ (long-dashed).}
   \label{eqazumithalvelgas}
\end{figure}

  Our initial models are motivated by the nuclear region of M100 (NGC 4321). 
We adopted a total gas mass of $\natd{4.7}{8} \, \msun$ distributed 
exponentially with a scale length of 300 pc within a radial range of 
\mbox{$R_{\rm in} = 30$ pc} to \mbox{$R_{\rm out} = 3$ kpc}. 
Thus, the central surface density of the gas is
about $750 \msunpctwo$. The small disk scale length
mimicks a central concentration of (molecular) gas observed in many galaxies.

The rotation curve was selected as a combination of a central rigid 
rotation and an outer flat rotation curve according to Eq.\ (\ref{eqvcirc}).
The velocity at infinity, $v_\infty$, was set to 178 km s$^{-1}$.
The transition parameter $n_t$ was selected to be 10, by this resulting in a
fairly sharp transition at the radius \mbox{$R_{\rm flat} = 100$ pc}. 
This rotation curve corresponds to a total dynamical mass 
$M_d(R) \sim v_c^2(R) R / G$ (including all components) 
of $\natd{7.3}{8} \msun$ within the central 100 pc.
In the region of rigid rotation the rotation period is about 
\mbox{$\natd{3.5}{6}$ yrs}. It increases outwards reaching
\mbox{$\natd{1.7}{7}$ yrs} at the half-mass radius of
the gaseous component at $R \approx 500$ pc.

According to the chosen rotation curve, mass profile and equation of
state (polytropic with $\gamma_g=5/3$), the minimum value of the Toomre
parameter is reached at a galactocentric distance of about 440 pc.
The constant in the equation of state (for the gaseous phase) was selected
to yield a minimum Toomre parameter of $Q_{\rm min} = 1.54$.
This choice corresponds to sound speeds between 4 and 11 km s$^{-1}$ 
within the central kpc (the higher value is reached in the center). 

\begin{figure*}
   \resizebox{\hsize}{!}{
     \includegraphics[width=17cm,angle=0]{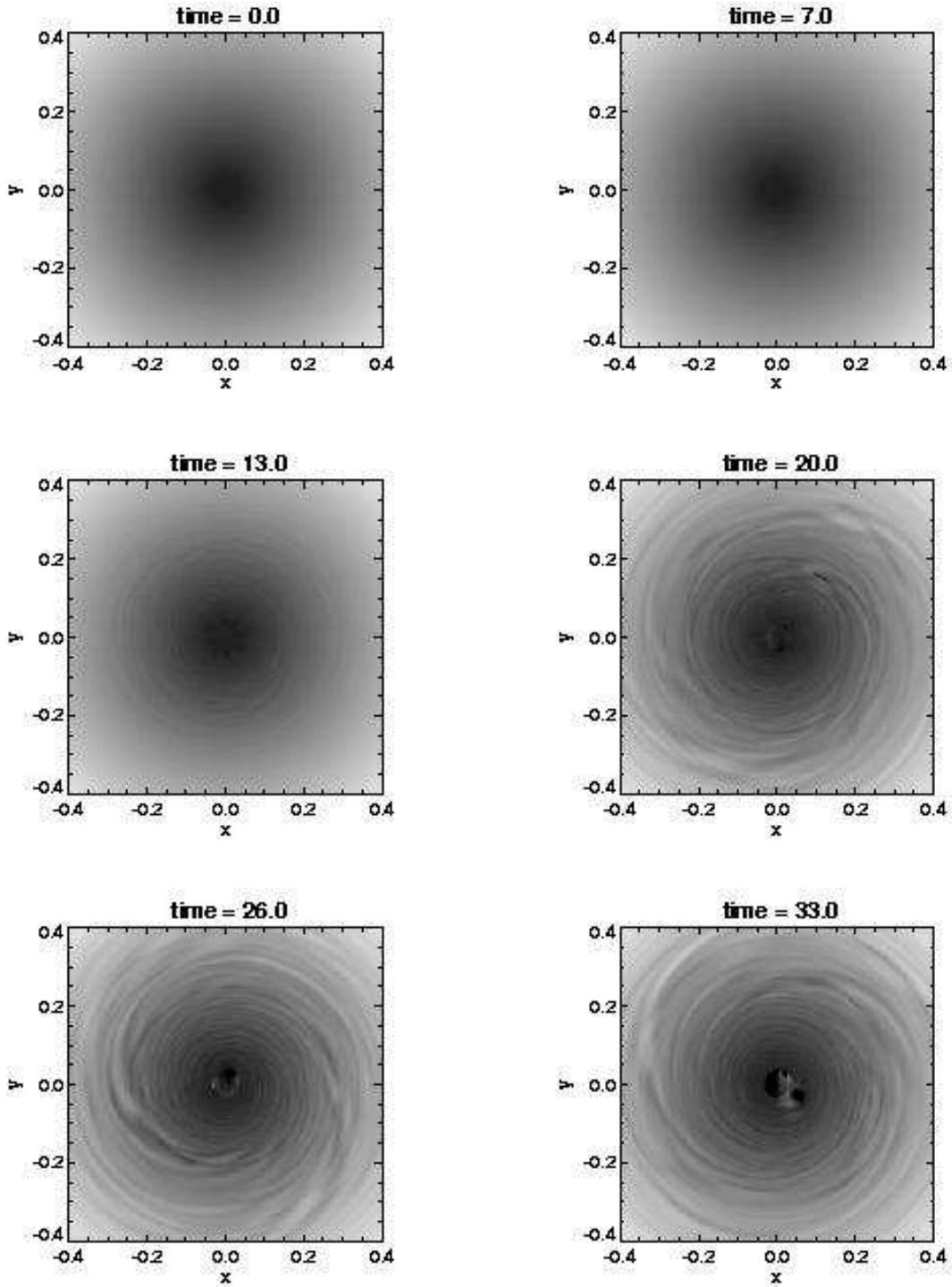}
   }
   \caption{Surface density of the gas component of the reference
    model at different times. A white area corresponds to surface
    densities of $10^2 \msunpctwo$ or lower, whereas black means
    $10^3 \msunpctwo$ or higher. The length unit is kpc and the
    time is given in $\natd{1.5}{7}$ yrs.}
   \label{m0037_surfden_gas}
\end{figure*}

\begin{figure*}
   \resizebox{\hsize}{!}{
     \includegraphics[width=17cm,angle=0]{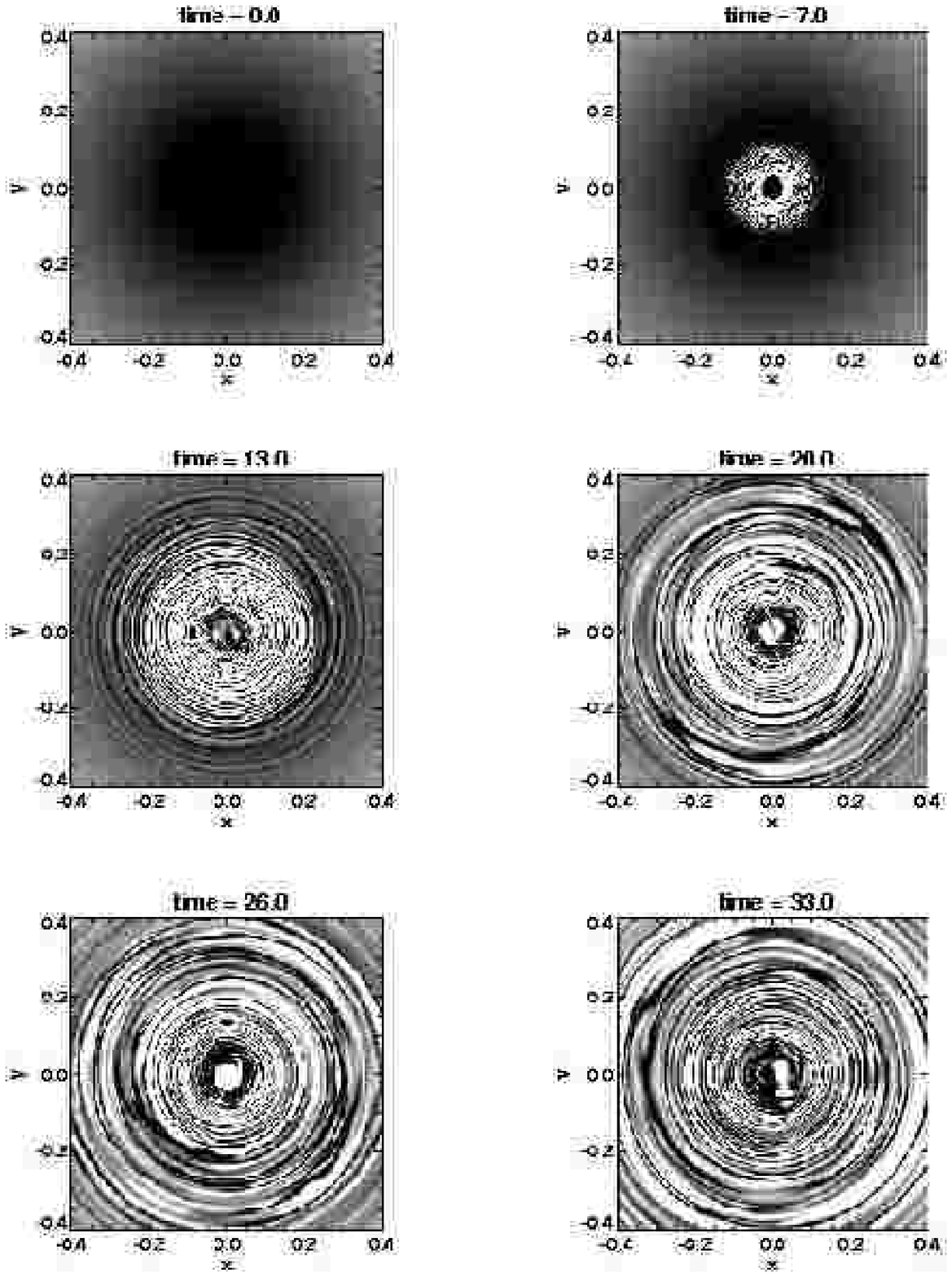}
   }
   \caption{Surface density of the dust component of the reference
    model at different times. A white area corresponds to surface
    densities of $1 \msunpctwo$ or lower, whereas black means
    $10 \msunpctwo$ or higher. The length unit is kpc and the
    time is given in $\natd{1.5}{7}$ yrs.}
   \label{m0037_surfden_dust}
\end{figure*}

The different contributions to the azimuthal velocity of the gas as well
as the rotation curve $v_c$ are shown in Fig.\ \ref{eqazumithalvelgas}:
The contribution $v_P$ of the pressure to the initial rotational equilibrium
velocity is very small. It reaches a maximum of \mbox{8.5 km s$^{-1}$} at a
distance of 450 pc from the center. Compared to the azimuthal speed 
$v_{\rm rot}$ of the gas $v_P$ is negligible as the almost identical values
of the azimuthal speed and the rotation curve demonstrate. The radial force 
attributed to the gravitational potential is dominated by the contribution 
of the dark halo plus stellar disk/bulge. The self-gravity of our live disk 
components (gas \& dust) gives rise to a maximum rotation velocity of
about 50 km s$^{-1}$. This value still exceeds the pressure contribution 
by far, whereas it is small compared to the ''halo'' contribution 
made out of stars and dark matter. Hence, the (gaseous) disk is submaximal 
by one order of magnitude. 

Assuming that the central region is dominated by stars (i.e.\ the dark 
matter mass fraction is small there), the velocity dispersion $\sigma_s$
of the stars must be a factor of 10 larger in order to form
a Toomre-stable maximum disk (their Toomre parameter scales
with $Q \sim \sigma_s / \Sigma$). Observations of the central stellar velocity
dispersions report such large values of about $\sigma_s \sim 100-150$ 
km s$^{-1}$ (e.g.\ H\'eraudeau \& Simien \cite{heraudeau98}). 
Hence even a maximum stellar disk would be as Toomre-stable as the
gaseous disk and more likely it would be dynamically hotter, i.e.\
more stable. On the other hand the scale height $h$ of such a 
stellar disk is fairly large. Estimating $h$ from $h = \sigma_s^2 / 
(\pi G \Sigma)$ yields for \mbox{$\sigma_s = 100$ km s$^{-1}$} and 
$\Sigma = 1000 \msunpctwo$ (the value of a maximum stellar disk at a
galactocentric distance of about 500 pc) a scale height of $h \sim 750$ pc. 
Hence, the stellar component is not flat at all within the central 
component. Together with the large Toomre parameter of the stellar
component this justifies the treatment of the stellar ''disk'' as a
''background'' potential instead of a dynamically live component.

In our reference model we set the dust mass to 2\% of the gas mass. 
This choice is close to the upper value reported for the average
dust-to-gas ratio in galaxies. On the other hand, the central regions
of galaxies are metal-enriched compared to the mean galactic values.
Taking the observed relation between the dust-to-gas ratio $r$ and 
the metallicity into account (e.g.\ Issa et al.\ \cite{issa90}), 
it is reasonable to assume a larger $r$ in the central regions 
than on average.

Since we treat the dust as a pressureless component, the initial
equilibrium azimuthal speed of the dust is equal to the 
circular speed $v_c$, whereas the azimuthal velocity of the gas
is slightly lower due to the pressure contribution to the radial
forces. The velocity difference between both components, however, is only
about 0.1-0.2 km s$^{-1}$ within the central kpc.

Though the choice of our initial model
was guided by observations of M100, it should be kept in mind that the 
following simulations are not meant to yield a model of 
M100\footnote{E.g.\ the weak bar is not included in our models.}, but to
study a ''typical'' central galactic region.

\subsubsection{Evolution of the reference model}

\begin{figure}
   \resizebox{\hsize}{!}{
     \includegraphics[angle=90]{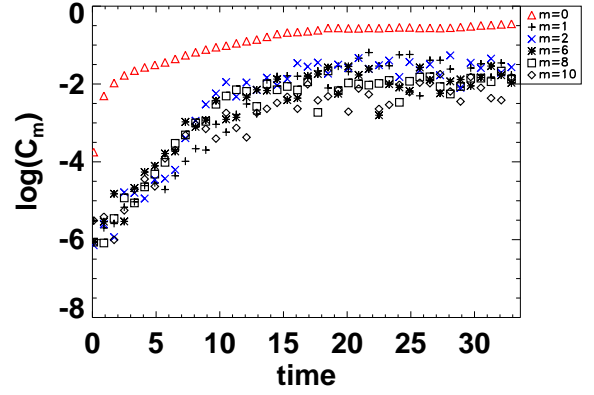}
   }
   \caption{Temporal evolution of the Fourier amplitudes of 
    the $m=0,1,2,6,8,10$-modes of the dust component in the
    reference model. The time unit is $\natd{1.5}{7}$ yrs.}
   \label{referencemodel_globalmodes_dust}
\end{figure}

\begin{figure}
   \resizebox{\hsize}{!}{
     \includegraphics[angle=90]{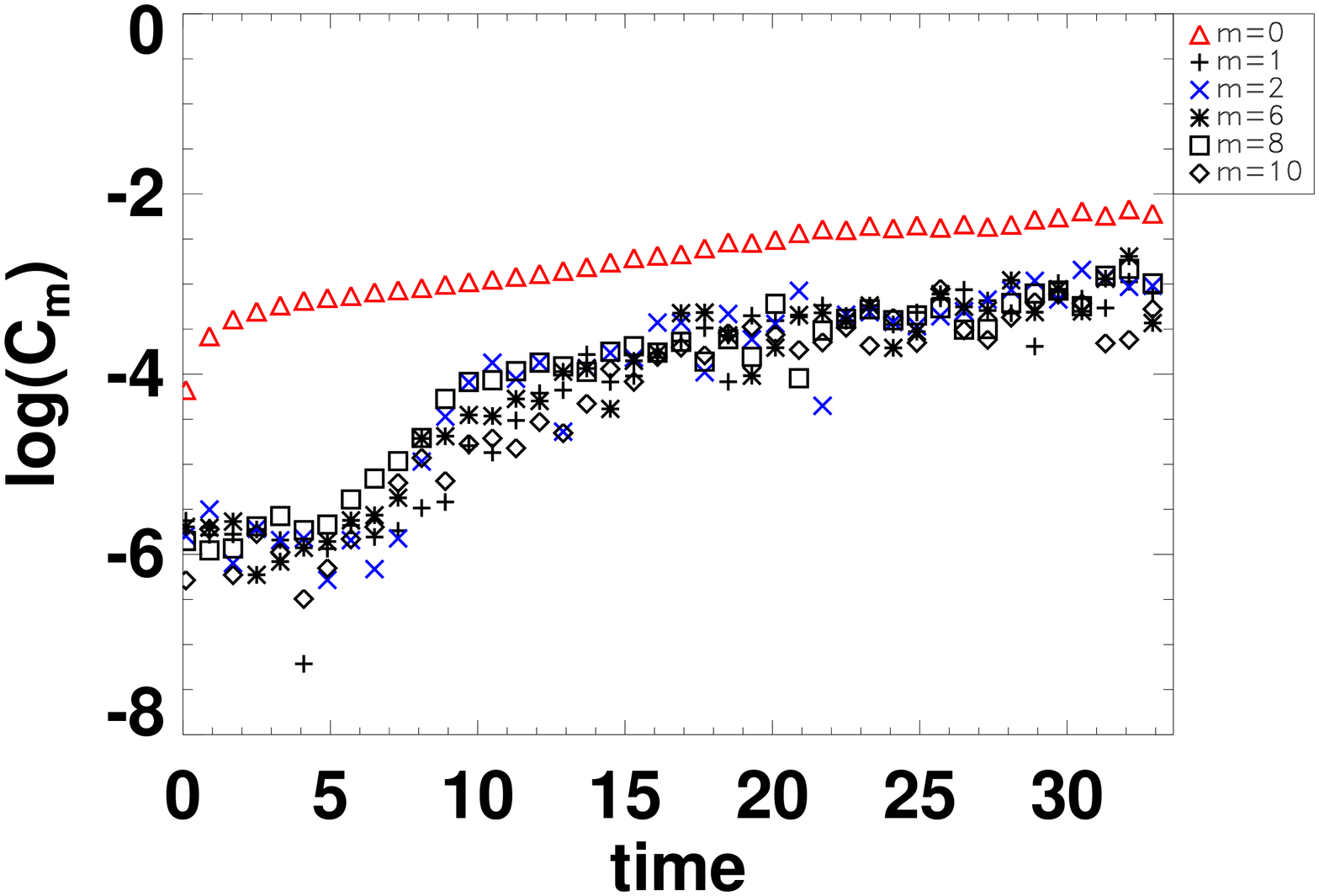}
   }
   \caption{Temporal evolution of the Fourier amplitudes of 
    the $m=0,1,2,6,8,10$-modes of the gas component in the
    reference model. The time unit is $\natd{1.5}{7}$ yrs.}
   \label{referencemodel_globalmodes_gas}
\end{figure}

\begin{figure}[tp]
  \resizebox{\hsize}{!}{
   \includegraphics[angle=90,width=8.5cm]{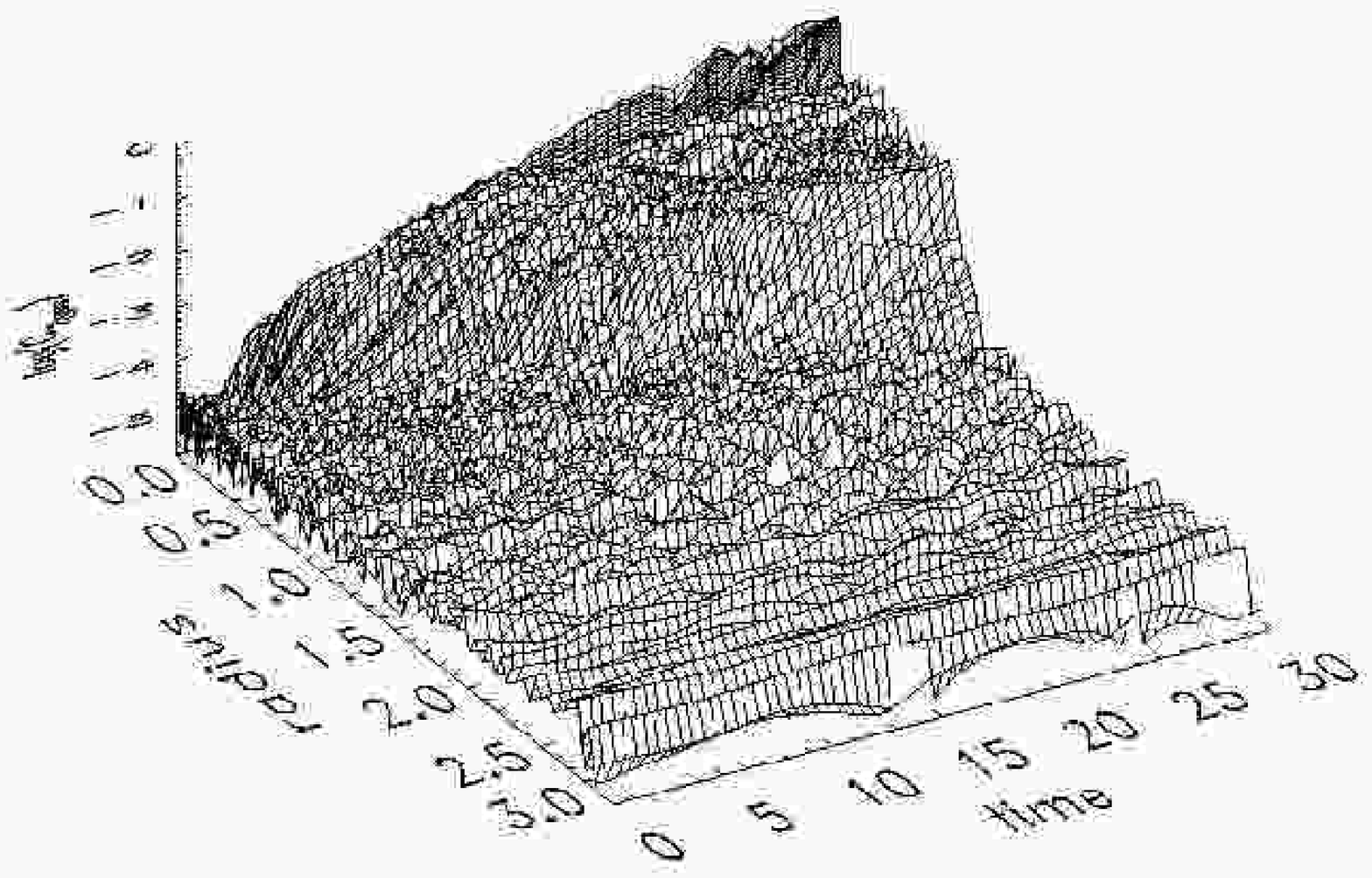}
   }
  \resizebox{\hsize}{!}{
   \includegraphics[angle=90,width=8.5cm]{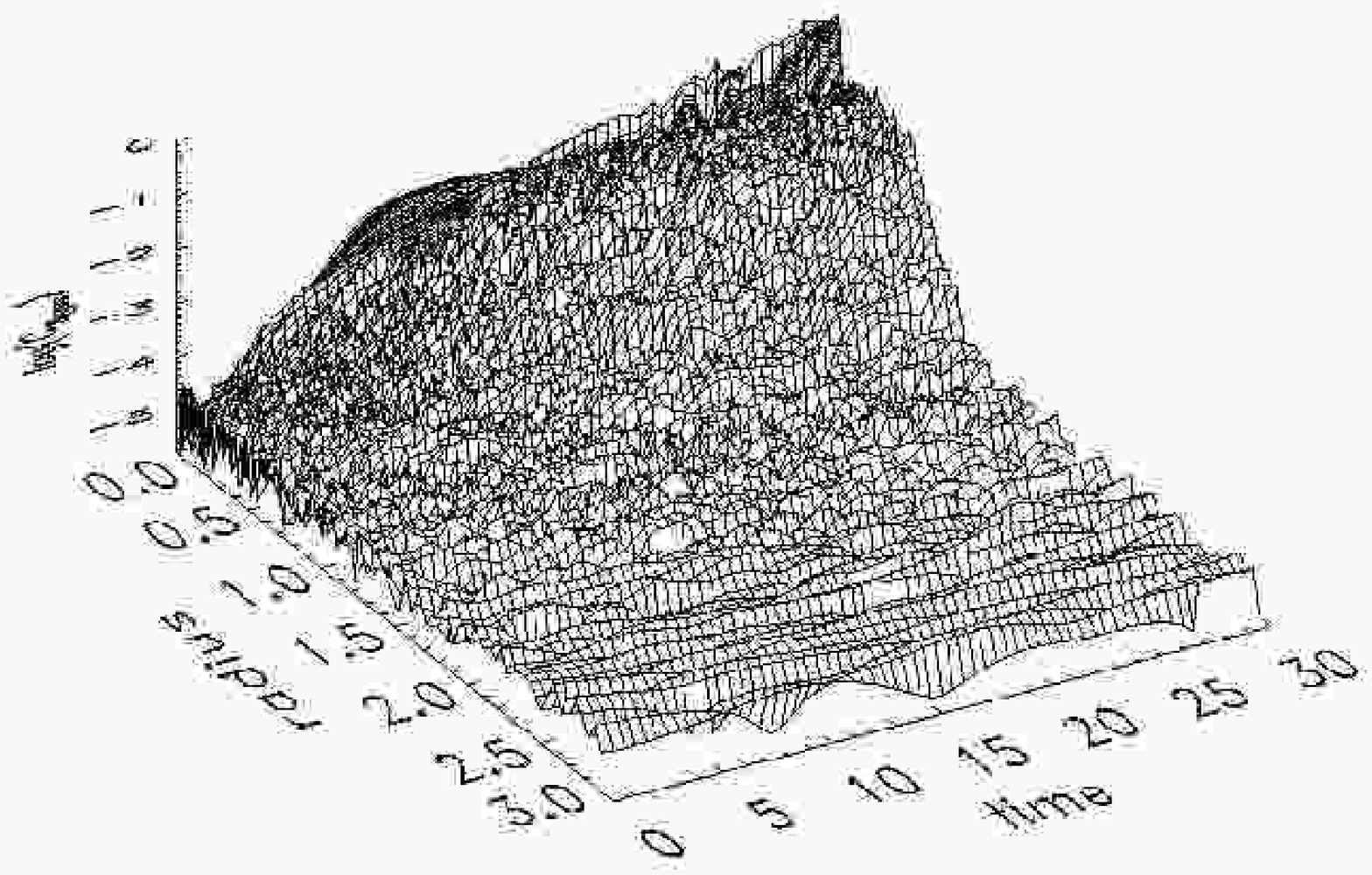}
   }
  \caption{Temporal evolution of the spatially resolved Fourier
    amplitudes for the $m$=2-mode (upper diagram) and the
    $m$=8-mode (lower diagram) for gaseous component of the reference model.
    The amplitudes are calculated according to Eq.\ (\ref{eqamplitude}) 
    integrating over small radial annuli defined by the numerical
    grid resolution. The time unit is $\natd{1.5}{7}$ yrs.}
   \label{referencemodel_spat_ampl_gas}
\end{figure}

   Fig.\ \ref{m0037_surfden_gas} shows images of the gaseous component
at different times. During the first 300 Myr ($t\sim 20 \approx 300$ Myr) 
only very 
weak features are discernible in the gas phase. Lateron, a multi-armed
and very patchy structure is formed which is basically concentrated to
the central 400 pc. After 390 Myr a small region close to the center
becomes highly unstable. Since we did not include any processes like
star formation or stellar feedback the growth of this clump is not
stopped. As a result the timestep in our simulations decreased to less
than a year. At this value we stopped the numerical simulation. 
The formation of such clumps, however, is a generic behaviour of our 
simulations. When we varied the initial conditions by using different
perturbations, clumps were still formed on a similar timescale 
($\sim 500$ Myr).

    In comparison to the gas, the dust becomes unstable at an earlier
time (Fig.\ \ref{m0037_surfden_dust}): already after 100 Myr ($t\sim7$)
structures are visible in the central 100 pc. These structures
are characterized by a large contrast, i.e.\ thin, but 
(relatively) dense regions, separated by broad areas devoid of almost
any dust. The positions of the dust peaks are strongly correlated with 
those of the gas, but sometimes with a small offset in the positions. 
The absolute values of 
those peaks are not well correlated with the corresponding peak values
of the gas density. This leads to a large scatter in the gas-to-dust 
mass ratios. Morphologically, the dust seems sometimes to be 
organized in rings instead of large-scale spirals like the gas.
Some of these rings are linked by dust lanes or small arcs.

  A more quantitative description of the structure formation is given
by the Fourier amplitudes. The dust component shows a fast exponential
growth of all components with a slight dominance of the high-$m$
modes (Fig.\ \ref{referencemodel_globalmodes_dust}). Different
to the single-component model discussed in Sect.\ \ref{singlecomponentmodel}
the dust modes begin to grow immediately at $t=0$. Their growth rates
are a factor of 2 larger than those found in the single-component model.
Additionally, the evolution is accompanied by a strong radial mass 
redistribution of the dust. After 150 Myr a saturation level of 
about $10^{-2}$ is reached.

  Though the dust has only 2\% of the mass of the gas, it destabilizes the
gaseous disk strongly (Fig.\ \ref{referencemodel_globalmodes_gas}). 
When dust is present, the modes of the gaseous phase commence to increase
already at about $t \sim 5 \approx 75$ Myr. In the single-component simulation
the gaseous disk has been stable until a time $t \sim 10 \approx 150$ Myr
(cf.\ Fig.\ \ref{singlecomponent_globalmodes}). This difference
becomes even more obvious when comparing the spatially resolved Fourier
amplitudes of the simulations with and without dust (Figs.\ 
\ref{singlecomponent_spat_ampl} and \ref{referencemodel_spat_ampl_gas}): 
the growing region is for both simulations restricted to the central area, 
but in the dusty model the growth sets in much earlier and with a 
larger growth rate.

Another difference to the single-component model concerns the dominant
modes. Whereas for the purely gaseous model the high-$m$ modes are
dominant, all modes grow more or less at the same rate
in the reference model. The superposition of all these modes  
with nearly equal amplitudes results in the patchy structure visible
in the spatial distribution of the components.

  The growth of the Fourier amplitudes of the gas shows four
stages (Fig.\ \ref{referencemodel_globalmodes_gas}).
In the first stage (until $t \sim 5 \approx 75$ Myr), 
the gas is almost unaffected
by the presence of dust and remains on its initial perturbation level.
In the second stage (until $t \sim 10 \approx 150$ Myr), 
the gas reacts on the grown instabilities in the dust, i.e.\ the 
formed very dense dust lanes, by 
becoming unstable, too. This phase of growth changes, when the dust reaches 
its saturation level. In the following third stage, the instabilities within
the gas are still growing but on a longer timescale. The last stage
is reached, when the gas phase saturates (which did not happen
until the end of this simulation).

The direct influence of the frictional force is visible when considering
the velocities. E.g.\ the residual azimuthal velocity, which is normalized to
the initial azimuthally-averaged equilibrium rotation speed, shows a small
but systematic deceleration of the dust component. 
Because the dust is treated as a pressureless phase, its rotation speed 
exceeds that of the gas. Therefore, the frictional force leads to a 
deceleration and, by this, to an azimuthal velocity which is 
0.05 km\,s$^{-1}$ smaller than the circular speed. The dust
gets out of rotational equilibrium resulting in a steady dust inflow.
This radial flow becomes larger with increasing coupling of the dust
to the gas. Hence, the $m=0$-Fourier amplitude (which
characterizes the radial mass reconfiguration) varies stronger when
increasing the coupling strength, while the higher Fourier amplitudes ($m>0$)
are less affected.

  The angular momentum removed from the dust is absorbed
by the gas which is accelerated in azimuthal direction. Due to
the large gas-to-dust ratio the acceleration is almost negligible and the
radial structure of the gas is much less affected. The specific angular 
momentum of the dust component decreases by 6.6\% during 450 Myr.
The overall relative conservation of angular momentum is about 
$\natd{2}{-6}$, i.e.\ the simulation is as accurate as the single-component
model. The total energy decreased by 0.8\% due to the dissipative nature
of the frictional force. Though this decrease is small, it exceeds by far
the intrinsic energy uncertainty of $\natd{5}{-5}$ found in the 
single-component simulation. Thus, the code is sufficiently
accurate to deal with the implemented dissipation rates.
Most of the energy dissipation takes place when the clumps are formed
and the system becomes highly non-linear.


\subsection{Parameter studies}
\label{parameterstudy}

   In a small set of parameter studies we investigated different aspects of
the dust's influence on nuclear gaseous disks. First, we compared 
in more detail the different dust-gas coupling schemes (Sect.\ 
\ref{frictionschemes}). Then we varied the gas-to-dust mass ratio
(Sect.\ \ref{gastodustratio}). In the sections \ref{toomrevariation} 
and \ref{eqofstate} we investigate the influence of the Toomre parameter 
and the equation of state of the gaseous component. Finally, different
minor aspects are summarized in Sect.\ \ref{miscellaneous}.


\subsubsection{Friction schemes}
\label{frictionschemes}

   In a first series of simulations we compared the different
coupling schemes Eqs.\ (\ref{eqatimescale}) and (\ref{eqadynamics})
applied to the reference model. Our simulations show that 
-- independent of the scheme applied for the dust-gas coupling -- 
the disks become more unstable when dust is present and the frictional 
timescale becomes not shorter than $10^4$ yrs.

\begin{figure}
   \resizebox{\hsize}{!}{
     \includegraphics[angle=270]{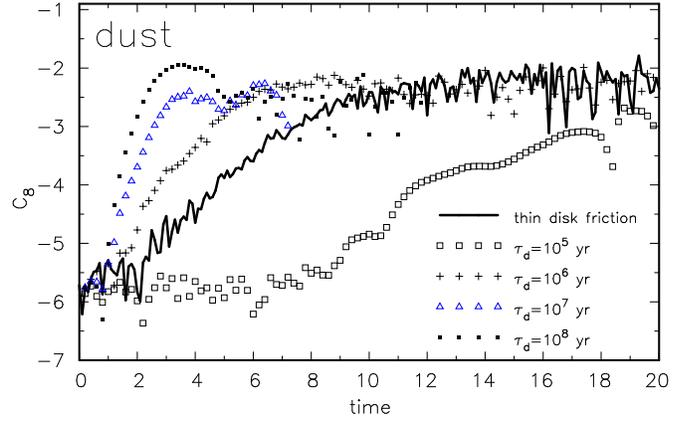}
   }
   \caption{Temporal evolution of the logarithmic Fourier amplitudes for 
    the $m=8$-mode of the dust component for different coupling
    timescales ($\tau_d=10^5$ yrs (open boxes), $\tau_d=10^6$ yrs (plus),  
    $\tau_d=10^7$ yrs (triangles) and $10^8$ yrs (filled boxes)) 
    and the reference model (thin disk friction; solid line). 
    The dust-to-gas mass fraction is 2\%. 
    The time unit is $\natd{1.5}{7}$ yrs.}
   \label{comp_dust_m8}
\end{figure}

In case of a constant frictional timescale $\tau_d$ 
according to Eq.\ (\ref{eqatimescale}) we investigated four values
of $\tau_d$: $10^8$, $10^7$, $10^6$ and $10^5$ yrs. In case of 
$\tau_d=10^8$ yrs the frictional timescale is of the order of the rotational
period at the outer boundary of the grid at 3 kpc. This timescale
exceeds the rotation period at the half-mass radius by a factor of 5.9
and the rotation period at the inner edge by a factor of 28.6.
The dominantly growing modes of the dust component are multi-armed modes
near $m=8$. The growth of the two-armed mode is delayed.
The dust modes grow very fast: already after 50 Myrs ($t \sim 3.5$) 
they reach their saturation level of about 1\%
(Fig.\ \ref{comp_dust_m8}). A weak coupling between gas and dust,
i.e.\ a large $\tau_d$, results in an almost decoupled dust disk which
is extremely unstable due to its lack of pressure support on small scales.

When $\tau_d$ is reduced by a factor of 100, the frictional timescale
is shorter than the dynamical timescale and gas and dust are strongly coupled.
In that case the growth of perturbations in the dust component is reduced, 
but still the dust evolves to saturation within the first $10^8$ years 
(Fig.\ \ref{comp_dust_m8}). The thin disk friction of the reference
model grows on an even longer timescale, corresponding to a frictional
timescale between $10^5$ and $10^6$ yrs. The model with the shortest
fixed friction timescale of $\tau_d=10^5$ yrs differs qualitatively
from the other simulations: whereas the other calculations exhibit an
immediate growth of modes, the $\tau_d=10^5$-model remains stable
for the first 120 Myr ($t \sim 8$) before it also develops instabilities.
The coupling to the dynamically warm gas becomes so strong that 
the growth of instabilities in the dust component is efficiently 
suppressed. Independent of these differences, the final saturation levels 
for all friction strengths are quite similar of the order of 1\%.

\begin{figure}
   \resizebox{\hsize}{!}{
     \includegraphics[angle=270]{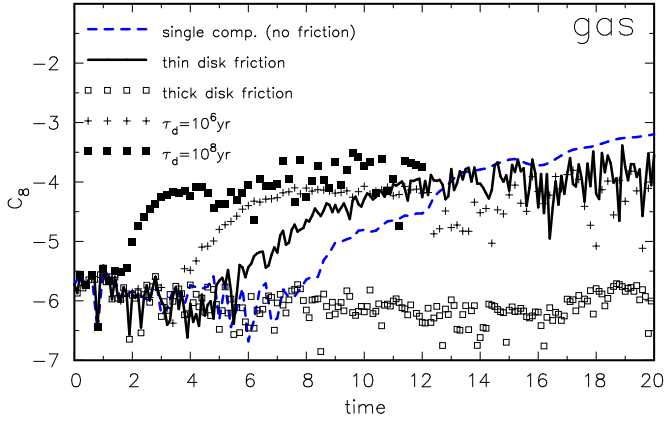}
   }
   \caption{Temporal evolution of the logarithmic Fourier amplitudes for 
    the $m=8$-mode of the gas component for different coupling
    timescales ($\tau_d=10^6$ yrs (plus) and $10^8$ yrs (filled boxes)), 
    the reference model (thin disk friction; solid line), 
    the single-component model (no friction; dashed line) and a model 
    with the ''thick disk friction'' (open boxes). The dust-to-gas
    mass fraction is 2\%. The time unit is $\natd{1.5}{7}$ yrs.}
   \label{comp_gas_m8}
\end{figure}

 In Fig.\ \ref{comp_gas_m8} the growth of the dominant $m=8$-modes
of the gaseous phase are shown for different coupling strengths. Similar
to the dust component, the evolution of the gaseous phase strongly
depends on the frictional timescale. In comparison to the single-component
model, friction strongly destabilizes the gaseous component, though its mass
is a factor of 50 larger than that of the dust. Also for the two more
realistic treatments of the dust friction, the thin and the thick disk limits,
the evolution of the gas phase deviates significantly from that of the 
single-component model. In the thin disk limit, the gas starts to form 
structures 25\% earlier than in the single-component model. If the
frictional timescale becomes even much shorter like in the ''thick disk'' 
limit, the gaseous disk is even stabilized by the dust (only after 
\mbox{$t=20 \sim 300$ Myrs}, the disk starts slowly to develop growing modes).
In that case, the friction acts like a very high viscosity which slows
down any growth of instabilities. 

  The more physical approaches for the dust treatment are characterized
by frictional timescales depending on the local gas and dust properties.
In case of a thin dust disk, the frictional timescale is given by 
Eq.\ (\ref{eqadynamics}), i.e.\ the dynamical timescale.
If we compare this ansatz with the simple approach of a constant
frictional timescale, we find that the thin-disk scheme is very similar
to a short frictional timescale between $10^5$ and $10^6$ yrs
(Fig.\ \ref{comp_dust_m8}), whereas the treatment according to
Eq.\ (\ref{eqacollision}) gives an even shorter timescale.

The coupling between the gas and dust component can be studied by 
comparing the evolution of the global modes of both components.
If one applies Eq.\ (\ref{eqacollision}) the coupling between gas and dust
is very strong and the modes have nearly identical temporal evolution.
If the frictional timescale is given by the dynamical time 
according to Eq.\ (\ref{eqadynamics}) the coupling
is less strong (see Figs.\ \ref{comp_dust_m8} and \ref{comp_gas_m8}).
The Fourier amplitudes of the dust exceed then those of the gas throughout 
the calculation. Especially in the phase of strongest growth, the dust is
clearly ahead of the gas. This indicates that the instability of the gaseous
disk is enhanced by the dust.


\subsubsection{Gas-to-dust ratio}
\label{gastodustratio}

\begin{figure}
   \resizebox{\hsize}{!}{
     \includegraphics[angle=270]{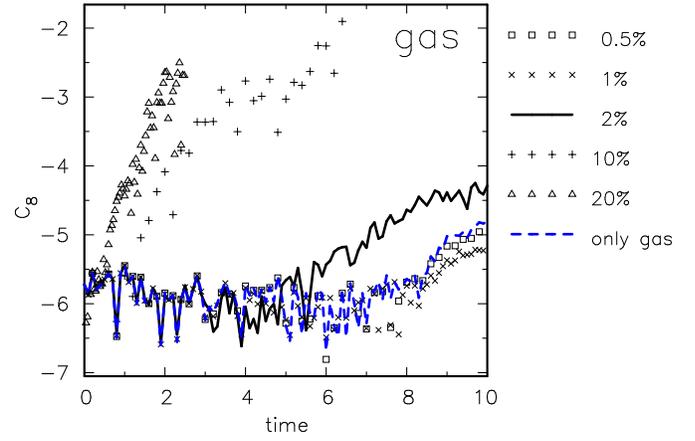}
   }
   \caption{Temporal evolution of the logarithmic Fourier amplitudes of the 
    dominant $m=8$-mode of the dust component for different
    dust-to-gas mass fractions: 0.5\% (open boxes), 1\% (crosses),
    2\% (reference model, solid line), 10\% (plus), 20\% (triangle)
    and the purely gaseous model (dashed line).
    The time unit is $\natd{1.5}{7}$ yrs.}
   \label{globalmodes_dusttogas_dust_m8}
\end{figure}

\begin{figure*}[tp]
  \centerline{\hbox{
  \includegraphics[angle=90,width=8.5cm]{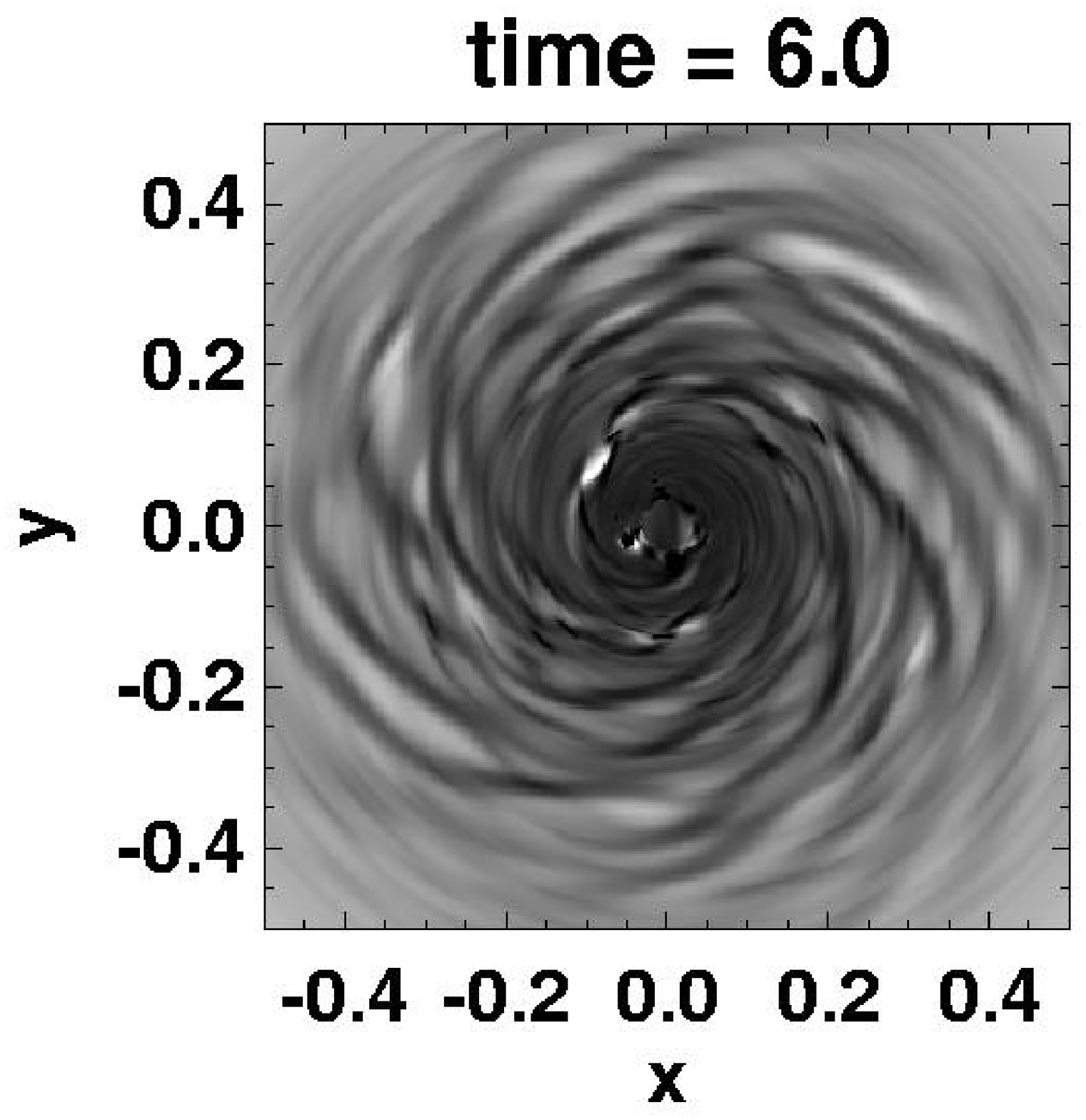}
  \includegraphics[angle=90,width=8.5cm]{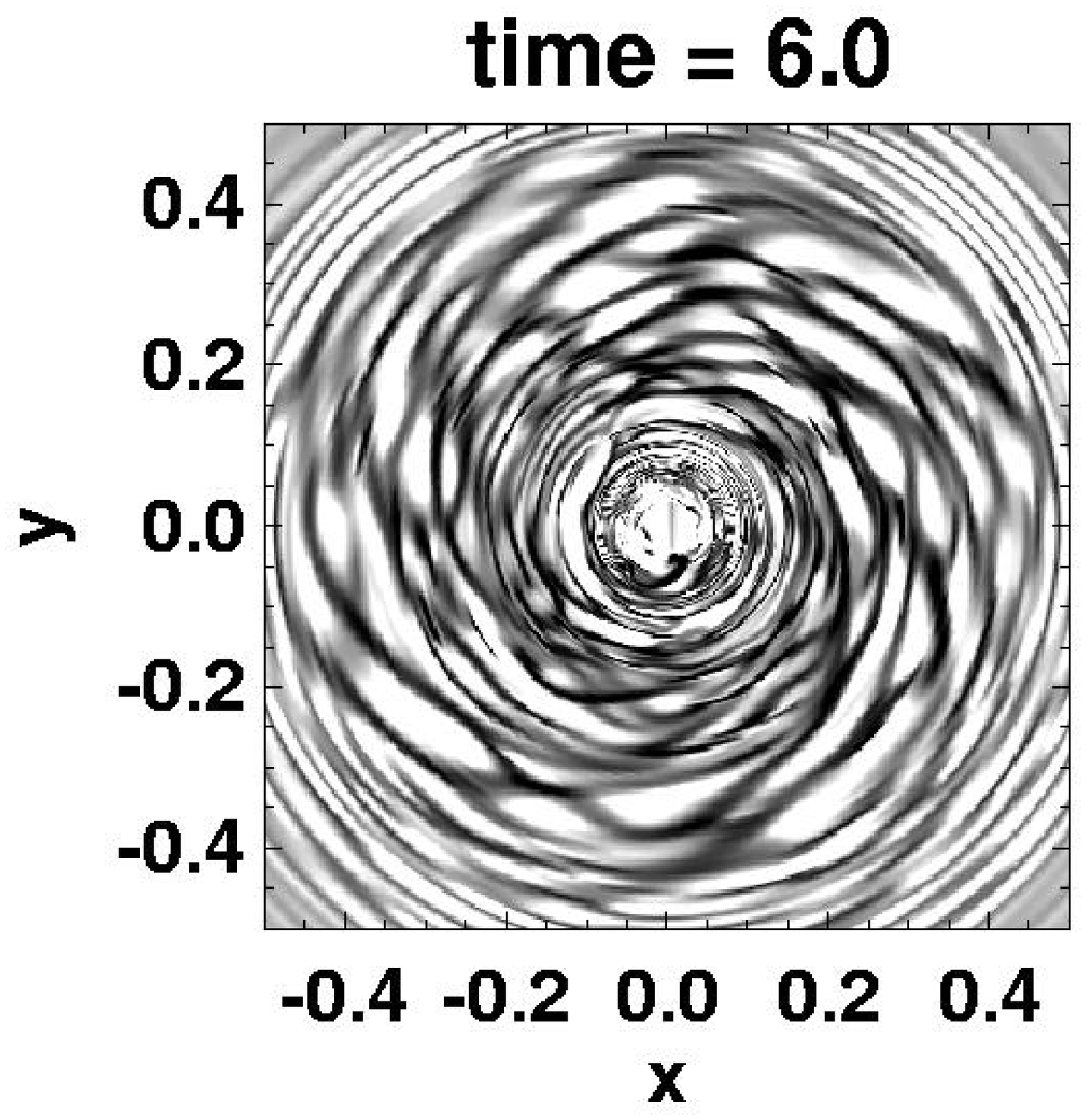}
  }}
  \centerline{\hbox{ 
  \includegraphics[angle=90,width=8.5cm]{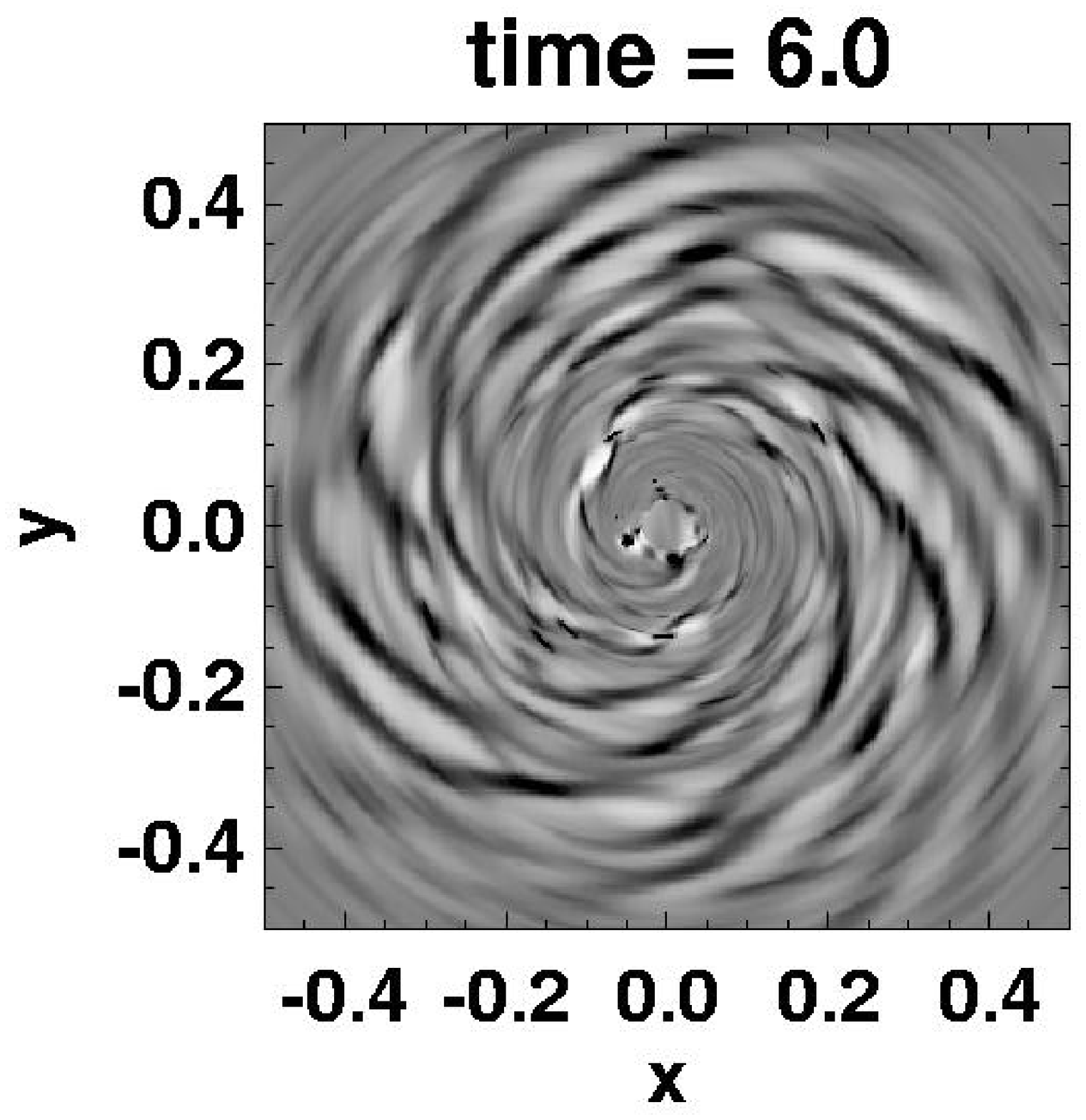}
  \includegraphics[angle=90,width=8.5cm]{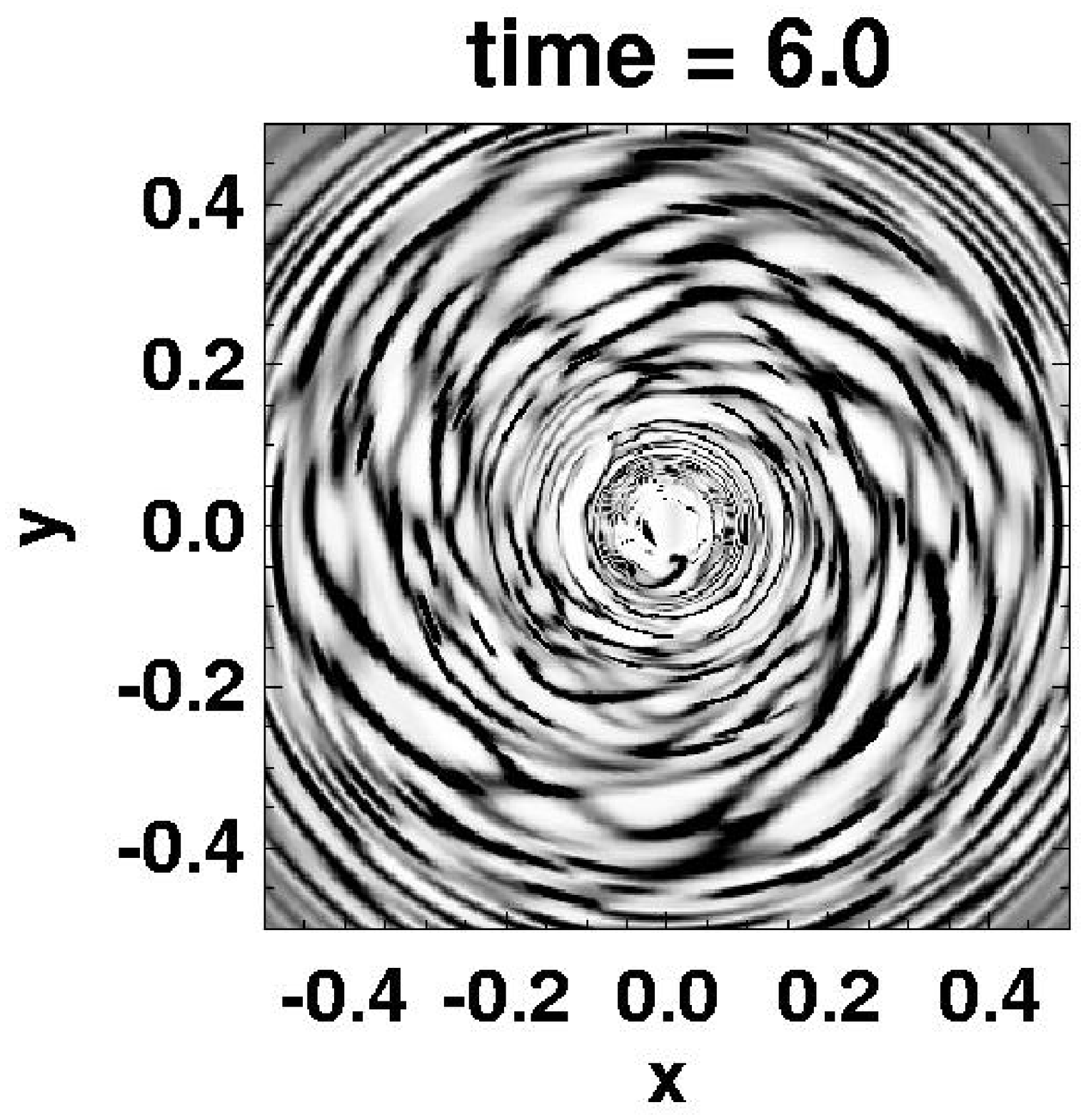}
  }}
  \caption{{\it Spatial distribution of the surface density and its
    perturbations (normalized to their initial values) of the
    gas and dust component after $t= 6 \sim 90$ Myr for a
    model with a 10\% dust mass fraction (relative to gas):
    $\Sigma_g$ (upper left; white: $10^{1.5} \msunpctwo$, black:
      $10^3 \msunpctwo$), 
    $\Sigma_d$ (upper right; white: $10^{0.5} \msunpctwo$, black:
      $10^2 \msunpctwo$), 
    $\Delta \Sigma_{g,d} / \Sigma_{g,d}$ for gas (lower left) and 
    dust (lower right). The latter are coded as follows:
    areas devoid of material are white, areas with a density enhancement 
    of a factor of 2 or more are black. The grey area seen at the outer 
    edges correspond to no deviation from the initial surface density.
   \label{muchdust_surface_denspert_image}}
   }
\end{figure*}

\begin{figure}
   \resizebox{\hsize}{!}{
     \includegraphics[angle=270]{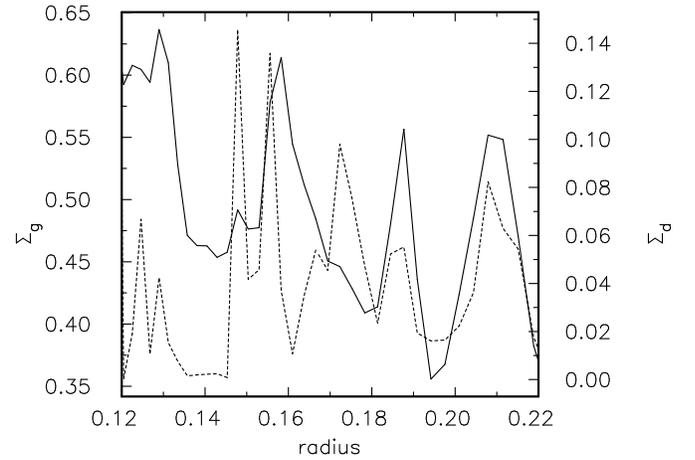}
   }
   \caption{Radial profile of the surface density of gas (solid) and 
    dust (dashed) along a radial line at $t=5$ for the 10\% dust
    fraction model. The surface densities are given in $10^3 \msunpctwo$, the
    radius in kpc.}
   \label{muchdust_surfrad_gasdust}
\end{figure}

  As a next step we varied the dust-to-gas mass ratio $r\equiv M_d/M_g$
from 0.5\% to 20\%. For all
these models the $m=8$-mode is dominant. A comparison of the corresponding
Fourier amplitudes with those of the purely gaseous model of Sect.\ 
\ref{singlecomponentmodel} shows that the critical dust-to-gas ratio 
$r_c$ for dust becoming dynamically unimportant is about 1\% 
(Fig.\ \ref{globalmodes_dusttogas_dust_m8}). For larger amounts of dust the 
destabilization of the gaseous phase becomes much stronger. For 
the reference model's value of $r=0.02$ the instability sets in 50 Myrs
earlier than in dustless or low-dust models while the modes remain on
their initial value for a latency period of 75 Myrs. Increasing $r$ to 10\%
reduces the latency time to almost zero. The growth rates increase by a
factor of 3-4 and the saturation level reached already after 30 Myrs is 
larger by at least one order of magnitude.

  Fig.\ \ref{muchdust_surface_denspert_image} shows the gas and dust
distribution of a model with a large dust-to-gas ratio of 10\%
in its saturation stage. The weak structures visible in the reference model 
become more pronounced as a comparison of the gas distributions shows 
(cf.\ also last images in Figs.\ \ref{m0037_surfden_gas} and 
\ref{m0037_surfden_dust}):
the patchy multi-armed structure is more emphasized due to larger
arm-interarm variations. Along the arms strong surface density
variations exist. Some arms are interrupted by low density areas. Many
spirals are not smoothly curved, but they show wiggles. Some arms seem
to merge with others. The dust distribution is highly correlated with
the gas distribution. However, the contrast between arm and interarm
regions is larger for the dust than for the gas. The surface
densities of the dust vary by about one order of magnitude, whereas
the contrast of the gas component is usually less than a factor of 2.
(Fig.\ \ref{muchdust_surfrad_gasdust}). 
The variations of the 
dust component along the arms are smaller than those of the gas. 
The structures formed in the dust are also thinner than those of the gas. 
Though there is a tight correlation between 
the positions of the maxima of the gas and dust phases, there is no 
correlation between their maximum amplitudes. The dust distribution is
characterized by a more cellular appearance compared to the spiral-like
morphology of the gas. Inspecting the density perturbations 
stresses this point (see lower diagrams in Fig.\ 
\ref{muchdust_surface_denspert_image}).

Another interesting aspect is that the dust is often located at the
boundaries around peaks of the gas distribution, preferentially at the
inner boundary. Examples are the dust peaks at $R=125$, 155 and
210 pc in Fig.\ \ref{muchdust_surfrad_gasdust}. Other dust peaks are
in regions with no or only weak gaseous density enhancements like those
at $R=150$ and 170 pc. And, of course, there are dust peaks at the same
locations as those of the gas, e.g.\ at $R=130$ and 190 pc. From that
variety of locations it is clear that there is no simple straightforward
correlation between the gas and dust mass distribution. Accordingly, one
expects large spatial variations of the dust-to-gas ratio.


\subsubsection{Toomre parameter}
\label{toomrevariation}

\begin{figure}
   \resizebox{\hsize}{!}{
     \includegraphics[angle=270]{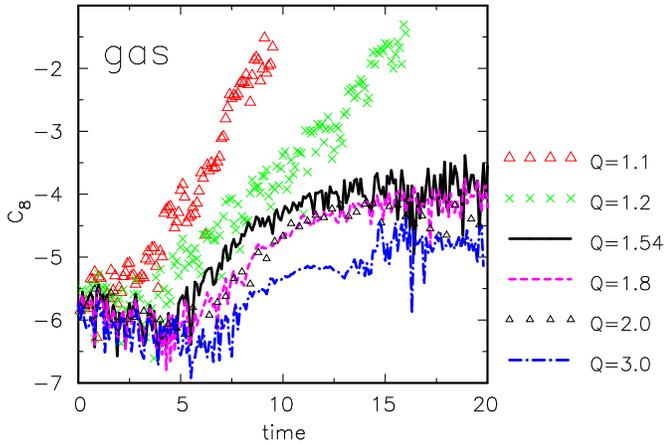}
   }
   \caption{Temporal evolution of the logarithmic Fourier amplitudes of
    the $m=8$-mode of the gas component for different initial (minimum)
    Toomre parameters of the disk: \mbox{$Q=1.1$} (triangle),
    \mbox{$Q=1.2$} (x), \mbox{$Q=1.54$} (reference, solid line),
    \mbox{$Q=1.8$} (dashed line), \mbox{$Q=2.0$} (boxes) and 
    \mbox{$Q=3.0$} (dot-dashed).
    The time unit is $\natd{1.5}{7}$ yrs.}
   \label{c8toomregas}
\end{figure}

\begin{figure}
   \resizebox{\hsize}{!}{
     \includegraphics[angle=270]{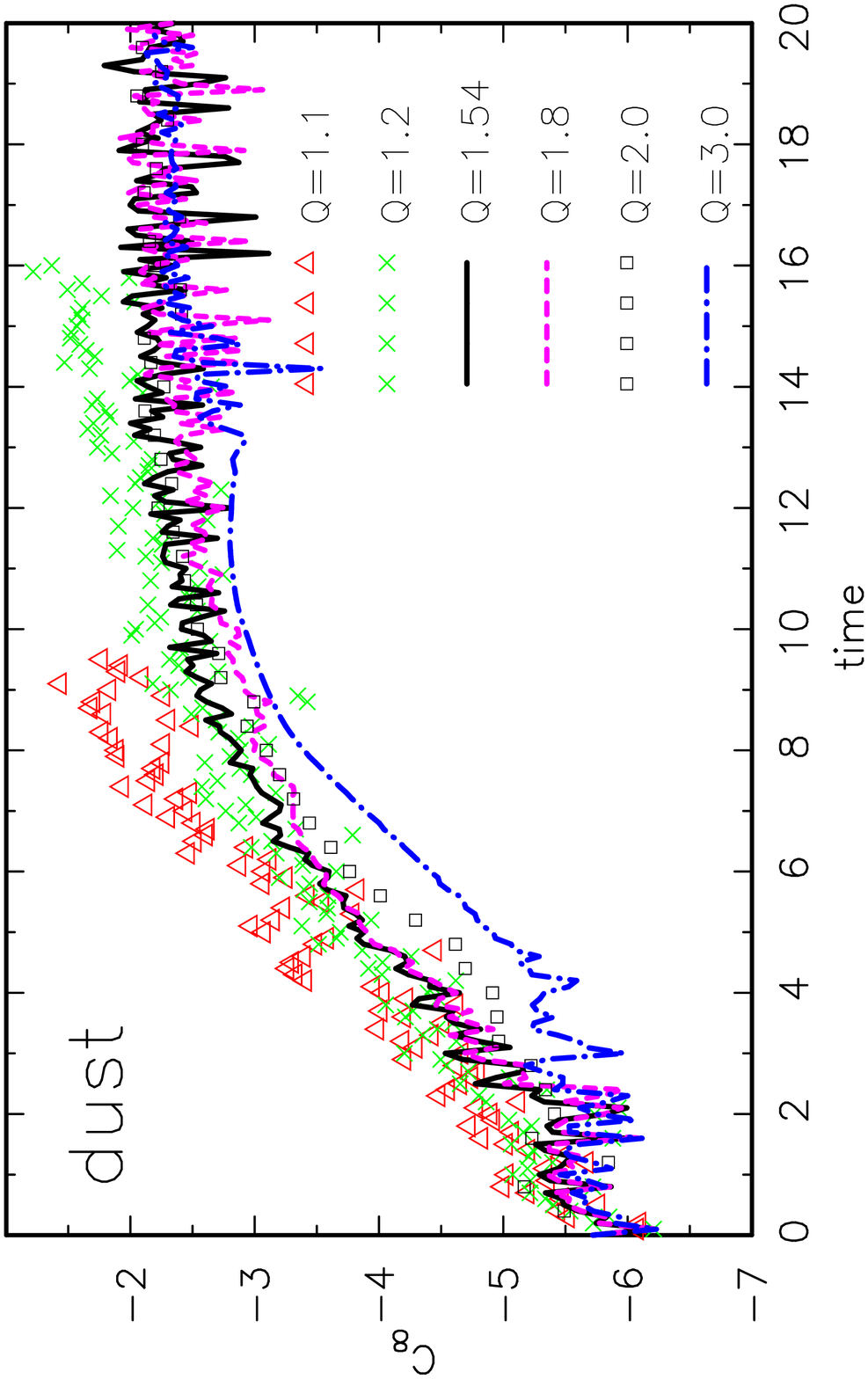}
   }
   \caption{Temporal evolution of the logarithmic Fourier amplitudes of
    the $m=8$-mode of the dust component for different initial (minimum)
    Toomre parameters of the disk: \mbox{$Q=1.1$} (triangle),
    \mbox{$Q=1.2$} (x), \mbox{$Q=1.54$} (reference, solid line),
    \mbox{$Q=1.8$} (dashed line), \mbox{$Q=2.0$} (boxes) and 
    \mbox{$Q=3.0$} (dot-dashed).
    The time unit is $\natd{1.5}{7}$ yrs.}
   \label{c8toomredust}
\end{figure}

\begin{figure*}[tp]
  \centerline{\hbox{
  \includegraphics[width=8.5cm]{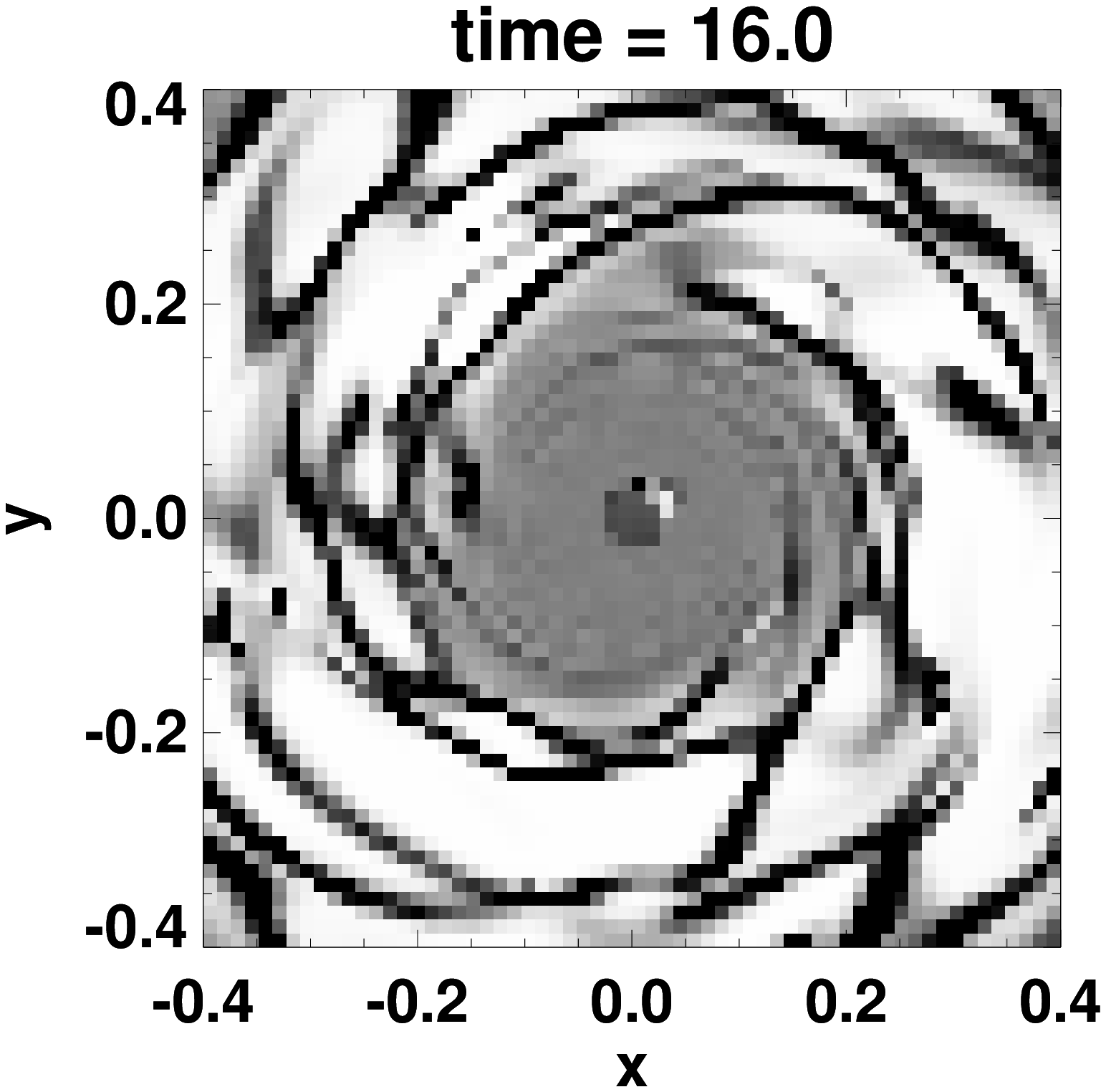}
  \includegraphics[width=8.5cm]{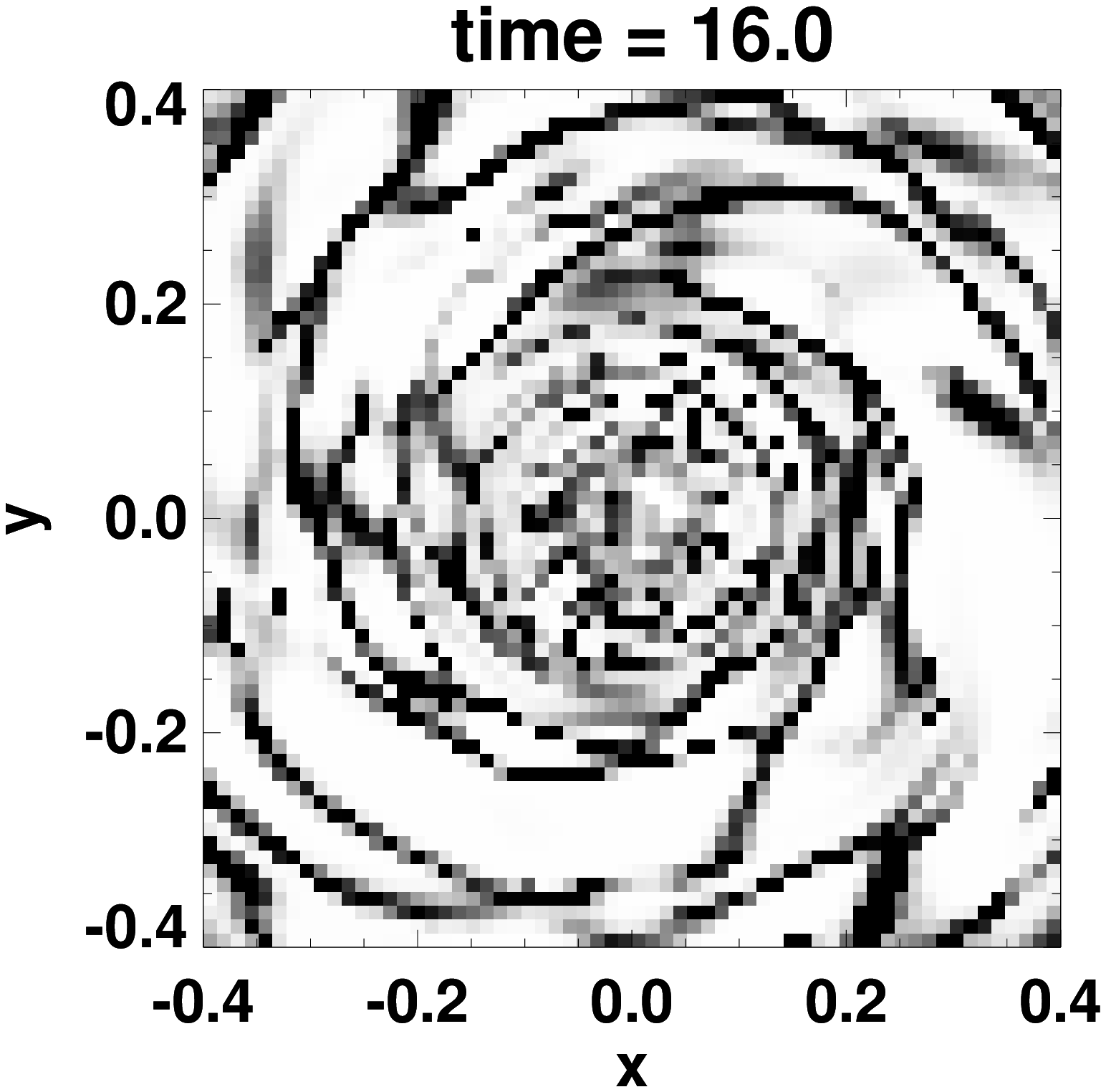}
  }}
  \caption{{\it Spatial distribution of the surface density perturbations
    (normalized to the their initial values) at \mbox{$t=16 \sim 240$ Myr}
    of the $Q=1.2$-disk: 
    gas component (left), dust (right).
    Areas devoid of material are white, areas with a density enhancement 
    of a factor of 2 or more are black. The grey area seen at the outer 
    edges correspond to no deviation from the initial surface density.
   \label{q12structure}}
   }
\end{figure*}

   The Toomre parameter usually turns out to be a key indicator for the 
stability of disks. Therefore, we performed a series of
simulations in which we studied dynamically cold, warm and hot disks, i.e.\
we varied $Q$ from 1.1 to 3.0. The dust-to-gas ratio was set to the value of
2\% identical to the reference model.
As expected from single-component simulations 
$Q$ has a strong impact on the structure formation of the gaseous phase 
(Fig.\ \ref{c8toomregas}). Dynamically cold disks ($Q$ close to unity)
are highly unstable, whereas hotter disks become more and more stable.
The saturation levels are much larger in case of cold disks (for which  
the non-linear regime is quickly entered). On the other hand, 
even for large $Q$ dusty gaseous disks are more unstable than single-component 
disks of the same $Q$ stressing the destabilizing influence of the dust.

   Qualitatively, this can be explained by the evolution of the dust
component whose growth rates are almost independent of the varied $Q$ values
(Fig.\ \ref{c8toomredust}). This is a consequence of the pressureless
evolution of the dust which leaves the dust disk almost unaffected by the
gaseous phase. There is just a small trend of the dust to become later
unstable with increasing $Q$. Additionally, the loss rate of specific 
angular momentum of the dust component by transfer to the gas varies by a 
factor of 7.3 when $Q$ is increased from 1.1 to 3.0. Both is caused by 
the larger pressure contribution to the rotational equilibrium 
of the gaseous phase in case of dynamically hotter systems. By this,
the gas rotates more slowly, if $Q$ is large, and the velocity difference
between gas and dust becomes larger, too. For small pressure contributions
to the azimuthal speed of the gas, the velocity difference between gas and 
dust scales like $\Delta v \sim \pder{P}{r} \sim K \sim Q^2$ ($K$ is the
constant in the equation of state Eq.\ (\ref{eqstate})). As a result the 
frictional force is also enhanced $\sim Q^2$ and the onset of instability 
of the dust component is slightly delayed.

   It is interesting to compare the spatial matter distribution of dust 
and gas in more detail. The simulation starting with a cold disk of
$Q_{\rm ini} = 1.2$ develops quickly a patchy multi-armed structure in the
gaseous phase. At the end of the simulation, when the dust reached
already the saturation stage and the gas is in the non-linear regime
of the growth of instabilities, the gaseous pattern is dominated 
by a multi-armed, irregular morphology outside 200 pc and a fairly regular,
mildly changed distribution in the central part (Fig.\ \ref{q12structure},
left). In contrast to the gas, the dust is also rather irregularly 
distributed in the circumnuclear region (Fig.\ \ref{q12structure}, right). 
Outside this nuclear region, but still within the central 200-400 pc prominent 
multi-armed spirals are visible. In that region the dust component
is strongly correlated with the gaseous phase. Moreover, in agreement
with the reference model the structures in the dust are more pronounced 
than those seen in the gaseous phase. The difference in the locations of 
unstable regions between both components becomes more pronounced at earlier
stages of the evolution, when the gas forms structures around a galactocentric
distance of 400 pc, whereas the dust becomes mainly unstable inside the 
central 100 pc. 

Both can be explained in terms of the local Toomre parameter $Q_{g,d}$ 
of the components: $Q_g$ takes its minimum value of 1.2 at
$R\sim 400$ pc growing to values of 1.5 at 200 pc. Therefore, the
gas becomes preferentially unstable in a broad ring at about 400 pc.
On the other hand, there is no pressure support at all for the dust component.
Hence, on small scales the dust is only prevented from collapse by
the friction with the gas (and, by this, partly due to the gas pressure).
On large scales, the dust (and the gas) can be stabilized by differential 
rotation. However, in the region of rigid rotation exists no differential
rotation and, therefore, the dust can form quickly fragments or rings.


\begin{figure}
   \resizebox{\hsize}{!}{
     \includegraphics[angle=270]{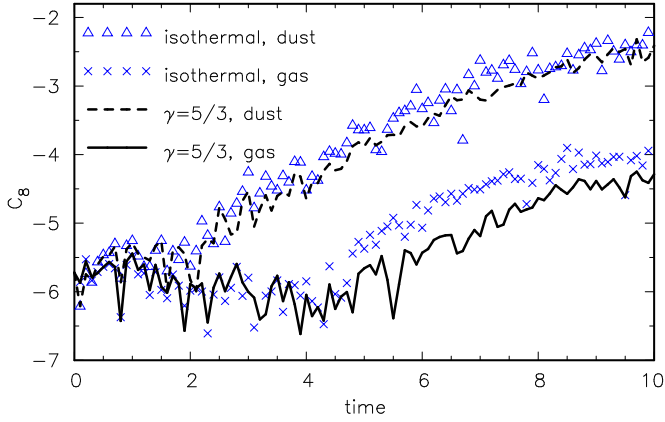}
   }
   \caption{Temporal evolution of the logarithmic Fourier amplitudes for 
    the $m=8$-mode of the gas and dust component for different equations
    of state (of the gas phase):
    $\gamma = 5/3$, reference model (gas: solid; dust: dashed) and 
    isothermal (gas: crosses; dust: triangles).
    The time unit is $\natd{1.5}{7}$ yrs.}
   \label{c8gasdustdifferenteqofstate}
\end{figure}

\subsubsection{Equation of state (gas component)}
\label{eqofstate}

  In order to investigate the influence of the equation of state
of the gaseous phase, we performed a simulation with an isothermal
equation of state ($\gamma = 1.0$) instead of $\gamma = 5/3$. The other
parameters were kept identical to the reference model. 
Both simulations give nearly the same results as the
growth rates of the dominant global amplitudes demonstrate
(Fig.\ \ref{c8gasdustdifferenteqofstate}). Thus, the behaviour of the 
dusty disks does not strongly depend on the adopted equation of state, provided
the minimum Toomre parameters are the same.


\subsubsection{Miscellaneous}
\label{miscellaneous}

   The influence of several ''technical'' parameters was studied by 
comparing the evolution of the dominant mode $m=8$ with that of a 
reference model.

{\it Grid boundaries.}
As a first test we investigated the influence of the radial extent of the 
grid by varying the inner and outer radial boundary by a factor of 0.5 
and 2, respectively. The growth rates, however, remained basically 
unaffected by the variation of the boundaries.

\begin{figure}
   \resizebox{\hsize}{!}{
     \includegraphics[angle=270]{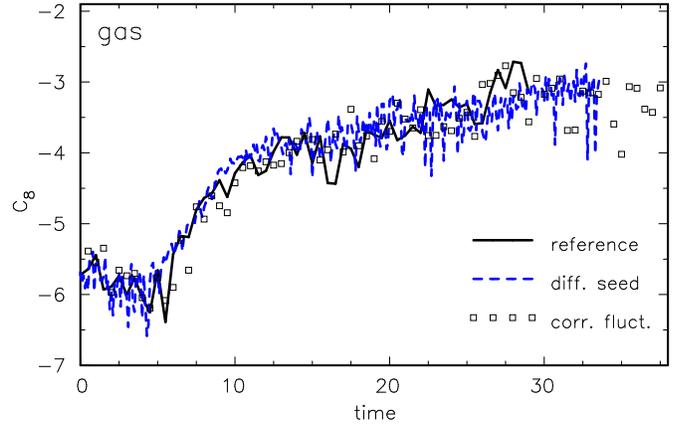}
   }
   \caption{Temporal evolution of the logarithmic Fourier amplitudes for 
    the $m=8$-mode of the gas component for different initial
    perturbations: reference model (solid), another set of random
    numbers, otherwise identical to reference model (dashed),
    locally identical relative overdensities for dust and gas
    (boxes). The time unit is $\natd{1.5}{7}$ yrs.}
   \label{c8gasdifferentseeds}
\end{figure}

{\it Grid resolution.}
  In order to test the influence of the grid resolution, we performed
a simulation with 686$\times$686 grid cells. Both simulations show only
minor differences, e.g.\ the evolution of the dominant mode $m=8$ is
rather similar in both cases for dust and gas: the instabilities 
set in at about the same time, the growth rates differ slightly, i.e.\
they are marginally larger in the high-resolution model. The saturation 
levels, however, are identical. Therefore, we conclude that 
our results are independent of the grid resolution.

{\it Different seeds.}
The influence of the initial setup of perturbations was tested in two ways:
First, a different set of random numbers was used to determine the initial
fluctuations. In a second simulation the initial perturbations of 
dust and gas were correlated by taking locally identical relative
perturbations, i.e.\ an overdensity of gas corresponds to the same
overdensity in dust (in the standard setup the perturbations were
uncorrelated). The global Fourier amplitudes were kept constant at a value
of $10^{-6}$. The dominant Fourier modes
evolve very similar in all these simulations (Fig.\ \ref{c8gasdifferentseeds}).
The main difference is the end of the simulations, which was chosen to be 
the time when the numerical timestep drops below 1 yr. In all simulations
this was initiated by the formation of a small scale gravitationally
unstable clump. In all three simulations this happened at a time
$t \sim 30 \approx 450$ Myr or later. Therefore, we conclude that
the formation of the main structures is rather insensitive to the 
detailed setup of initial perturbations.

\begin{figure}
   \resizebox{\hsize}{!}{
     \includegraphics[angle=270]{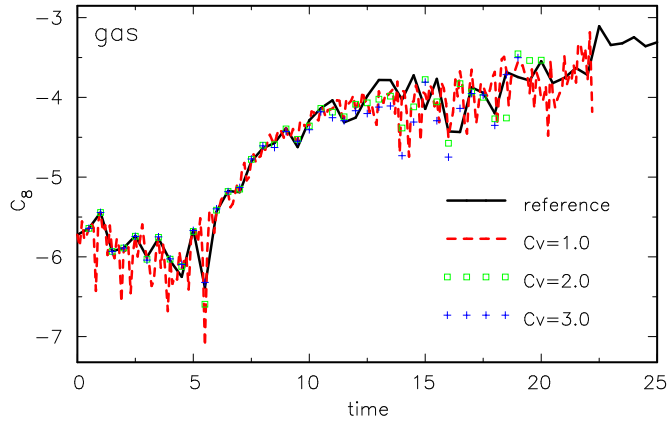}
   }
   \caption{Temporal evolution of the logarithmic Fourier amplitudes for 
    the $m=8$-mode of the gas component for different artificial
    viscosity: reference model: no artificial viscosity (solid), 
    $C_{\rm vis} = 1.0$ (dashed), $2.0$ (boxes), $3.0$ (plus).
    The time unit is $\natd{1.5}{7}$ yrs.}
   \label{c8gasdifferentartvis}
\end{figure}

{\it Artificial viscosity.}
The use of artificial viscosity is a standard tool in hydrodynamics
in order to cope with shocks. Principally, shocks
are expected in the late stage of the evolution, especially during the 
non-linear or the saturation stage. On the other hand, artificial viscosity
is formulated qualitatively similar to the terms describing friction,
i.e.\ it couples to velocity differences. Therefore,
we did not apply artificial viscosity to the models shown so far
in order to avoid a spurious mixing of effects induced by friction 
and artificial viscosity.

   However, we wanted to test the possible impact of (artifical) viscosity
on our results. Therefore, we implemented the von Neumann-Richtmyer 
artificial viscosity described in Stone \& Norman (\cite{stone92}) and 
modified the determination of the timestep in our code accordingly. 
The free constant $C_{\rm vis}$ (called $C_2$ in Stone \& Norman) 
describes roughly the number of cells over which a shock is smeared out. 
$C_{\rm vis}$ was varied from 1 to 3. The evolution of the global modes
was basically unaffected by the application of artificial
viscosity as the similar Fourier amplitudes in Fig.\ 
\ref{c8gasdifferentartvis} demonstrate. Hence, shocks seem to be less
important for the structures formed in our simulations. This, however,
might change when strongly barred nuclei are investigated as the
simulations by Athanassoula (\cite{athanassoula92}) have revealed.

\begin{figure}
   \resizebox{\hsize}{!}{
     \includegraphics[angle=270]{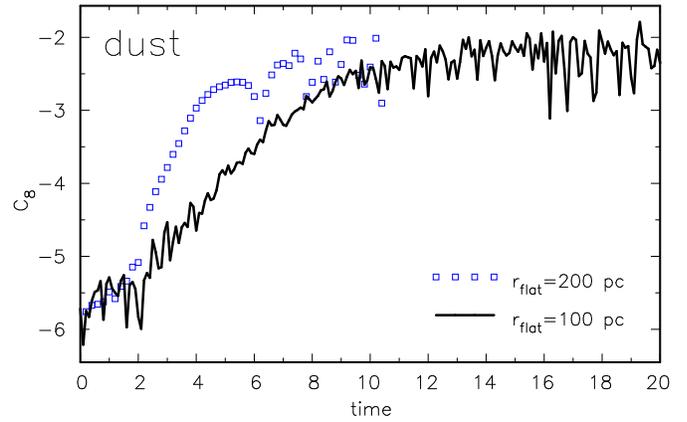}
   }
   \caption{Temporal evolution of the logarithmic Fourier amplitudes of
    the \mbox{$m=8$}-mode of the dust component for different transition
    positions $R_{\rm flat}$ of the rotation curve:
    reference model, \mbox{$R_{\rm flat} = 100$ pc} (solid) and 
    \mbox{$R_{\rm flat} = 200$ pc} (open boxes).
    The time unit is $\natd{1.5}{7}$ yrs.}
   \label{c8dustdifferentrflat}
\end{figure}

{\it Transition to flat rotation.}
In the reference model the region which became unstable first
was given by the transition from rigid rotation to the flat part of the
rotation curve, i.e.\ at a distance of \mbox{$R_{\rm flat} = 100$ pc}
(cf.\ also time $t=7$ in Fig.\ \ref{m0037_surfden_dust}). Within that
region the circular frequency is constant which means that a perturbation
always acts on the same neighbouring gas packets and differential
rotation can not pull apart those gas elements, by this preventing
gravitational instability. Contrary to that, in the flat part of the
rotation curve differential rotation stabilizes additionally the disk.

Fig.\ \ref{c8dustdifferentrflat} compares the $m=8$-modes for a
rotation curve with a transition to the flat part
at \mbox{$R_{\rm flat} = 200$ pc} instead of 100 pc as in the reference
model. In agreement with the simple considerations of the previous paragraph 
the dominant mode grows faster: after a short initial
period the region of instability extends beyond 100 pc making a larger
area unstable, by this enhancing the growth rate of the $m=8$-mode.
Compared to other modes the dominance of the high-$m$ modes is even 
more pronounced, because they affect a larger area and, thus, more mass.
The saturation levels, however, are nearly identical in both simulation
which can be understood as a consequence of the only small mass fraction
($<10\%$) which resides in the area between 100 and 200 pc.


\section{Discussion}
\label{discussion}


\subsection{Preliminary remarks}
\label{preliminaryremarks}

   From previous works it is well known analytically as well as numerically
that adding a cold component can destabilize galactic disks 
(e.g.\ Jog \& Solomon \cite{jog84}; Orlova \etal \cite{orlova02}). 
In our example the cold component is the dust phase, 
whereas the dynamically warm or hot component is the gas. A quick (but dirty) 
first estimate of the stability of the components can be done by inspecting
the Toomre stability parameter, i.e.\ stability is provided if $Q$ 
exceeds unity:
\begin{equation}
   Q \equiv \frac{\sigma \kappa}{\pi G \Sigma} > 1 \,\,\, .
\end{equation}
$\sigma$ is here the velocity dispersion of the component of interest, e.g.\
the sound velocity in case of a gaseous phase.
This criterion is exact only for axisymmetric perturbations in flat, 
single-component gaseous disks. Qualitatively, however, it holds for
a large variety of systems, if the RHS is replaced by a slightly larger\
numerical value. E.g.\ in case of stars one has a value of $3.36/\pi$
(BT87) or for non-axisymmetric modes one gets in case of flat rotation
curves $\sqrt{3}$ (Polyachenko \etal \cite{polyachenko97}).

   For a two component system the Toomre criterion gives only a good
estimate of the stability in some limiting cases in which the dynamics
of the components can be well separated, i.e.\ the meaning of the dynamically
active mass or surface density $\Sigma$ and its stabilizing pressure or
velocity dispersion $\sigma$ are well defined. This situation holds
for some of the systems under investigation here, e.g.\ they can be
described as cold dust disks embedded in a dynamically warm or hot gaseous 
disk which is in rotational equilibrium. 
The Toomre parameter is then determined by the external rotation curve,
the dust distribution $\Sigma=\Sigma_d$ and its velocity dispersion.
Since the dust is assumed to be pressureless, i.e.\ $\sigma=0$
the Toomre parameter vanishes. Thus, without other stabilizing processes
the dust disk would become gravitationally unstable to all perturbations 
with short wavelengths\footnote{The same result can be derived from a 
more detailed stability analysis of two-component systems 
(Jog \& Solomon \cite{jog84}).}.

  In case of a weak coupling between gas and dust, the dust disk would
evolve like a two-component disk. Especially the dust disk would become 
violently Jeans-unstable, provided the dust remains
pressureless as assumed by Noh \etal (\cite{noh91}). The latter would be the
case, if the increase of velocity dispersion of the dust phase
can be compensated by dissipative processes like inelastic collisions of dust 
particles. As a result the dust might form dense regions, eventually
dominated by dust. However, there is no evidence for the existence of
such pure ''dust balls'' which means that there is either a sufficiently 
strong coupling between dust and gas or there is no efficient dust ''cooling'' 
(or both of them). In the case of protoplanetary disks Weidenschilling 
(\cite{weidenschilling80}) and Cuzzi \etal (\cite{cuzzi93}) stressed
the importance of turbulence which limits the density in the dust layer, 
by this preventing the formation of dust-dominated regions. Similarly,
small-scale turbulence in galactic disks might be an important (dynamical) 
heating mechanism, even in the absence of a drag force. On the other hand,
in the limiting case of a very short frictional timescale
the velocity of the dominant gaseous phase is simply imprinted on the 
dust component. The behaviour of the complete system is then again well 
described by that of a single component system characterized by the gas 
parameters.

  The most interesting regime is that where the coupling between gas
and dust is intermediate and the dominant gaseous component is 
dynamically not too hot, i.e.\ not all instabilities of the gaseous 
component are completely suppressed. 


\subsection{Is the dust component dynamically important?}

   The simulations have revealed that the dust component is dynamically 
important only, if the dust-to-gas mass ratio $r$ exceeds 1\%. This
value is close to values of $r$ reported for the interstellar
dust surrounding the solar system. In-situ measurements of the
gas-to-dust ratio performed with the satellites Ulysses and Galileo
in the heliosphere give a value of $r \sim 1/94$ (Frisch \etal 
\cite{frisch99}). If one corrects their sample (which was dominated by 
micron-sized grains) for smaller grains missed because of the interaction
with the heliosphere $r$ increases by a factor of 2 to
$\sim$ 2\%, the value we used for our reference model. 

  On the other hand, there are several dust-to-gas determinations 
from interstellar absorption lines towards different directions (e.g.\ 
\mbox{$\epsilon$ CMa}) which show a large scatter of $r$ giving values 
down to only  $r \sim 0.2-0.3\%$ (Frisch \etal \cite{frisch99}). 
Though the exact numbers strongly depend on details of the analysis 
(e.g.\ the mass distribution of the grains, the gas column densities, 
the adopted stellar abundances), the high local variability seems to be 
characteristic for the dust distribution. 

  In external galaxies measurements of the dust-to-gas mass ratio
are derived from IRAS observations. Young \& Scoville (\cite{young91})
give a mean value of $r \sim 1/600$. They explain the large discrepancy to the
Milky Way ($\sim$ 1\%) by the assumption that the bulk of dust might be cold
($T<30$ K) radiating at wavelengths larger than 100 $\mu$m,
whereas IRAS is mainly sensitive for ''warmer'' dust. Detailed
determinations of the dust-to-gas mass ratio also suffer from 
difficult gas mass estimates. Especially, in the central
regions where molecular hydrogen is often the dominant gas phase, the
gaseous mass can only be inferred from CO measurements and a subsequent 
conversion to the H$_2$ mass. Uncertainties in this conversion might 
easily accommodate a factor of two in the mass determinations.

Though there are only a few dust-to-gas mass determinations, there is
a clear close-to-linear correlation of $r$ with the metallicity 
(e.g.\ Issa et al.\ \cite{issa90}, Hirashita \cite{hirashita99}). 
In case of M51 the dust-to-gas ratio is about 2\%, whereas for dwarf 
galaxies like the Magellanic Clouds $r$ is a factor of ten or more smaller.
Edmunds \& Eales (\cite{edmunds98}) estimated an upper total dust mass
limit for spiral galaxies with a standard yield of $p=0.01$ to be
0.2\% of the total baryonic mass fraction. Assuming a gas mass fraction
of 10\%, this gives an upper dust-to-gas ratio of about 2\% averaged
over the whole galaxy. In the central regions of galaxies, however,
the effective yields are larger, e.g.\ Pagel (\cite{pagel87}) gives a value
4.5 times larger than in the disk, and the limits on $r$ are accordingly
higher. Therefore, dust-to-gas mass ratios of 1\% or slightly more
seem not to be unusual high for normal spiral galaxies.

  Due to all these uncertainties, however, it is difficult to give a 
reliable global value for the dust-to-gas mass ratio in galactic centers. 
Local measurements yield
dust-to-gas ratios exceeding the critical 1\% margin, whereas
global ones favour smaller values. However, both estimates might suffer
from systematic effects (and the differences of both are a hint for
that). Therefore, it seems that dust-to-gas mass ratios of 1\% or more
for normal galaxies cannot be ruled out from current data. 
Especially in the strongly obscured metal-enriched nuclear regions 
the dust mass fractions might exceed the locally determined values.

It is also interesting to note that a strong place-to-place variation 
of the dust-to-gas ratio is found in our simulations. This is 
especially remarkable, because we started with an almost uniform 
dust-to-gas ratio (deviations of the order $10^{-6}$). Hence, even without
an inhomogeneous dust production, large spatial gas-to-dust 
variations are produced.


\begin{figure}
   \resizebox{\hsize}{!}{
     \includegraphics[angle=270]{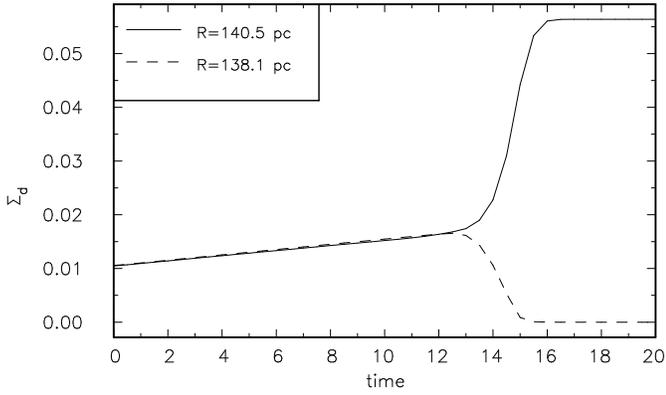}
   }
   \caption{Temporal evolution of the azimuthally averaged
    surface density of the dust component at two neighbouring radial
    cells ($R\sim 140$ pc) for the hot gaseous disk model with 
    $Q_{\rm min}=3.0$. The time unit is $\natd{1.5}{7}$ yrs, the surface 
    densities are given in $10^3 \msunpctwo$.}
   \label{surfavgdust_time}
\end{figure}

\subsection{The friction instability}
\label{frictioninstability}

  An interesting difference between the single-component model and
the dusty disks is that the dust forms ring-like perturbations
in the central area, whereas the gas develops initially only high-$m$ spiral 
modes (compare e.g.\ Figs.\ \ref{singlecomponent_denspert_image} and
\ref{m0037_surfden_dust}). Obviously, the dust introduces an additional
axisymmetric instability as the evolution of the azimuthally averaged
surface densities of the dust component show (Fig.\ \ref{surfavgdust_time}).

An axisymmetric instability due to a drag force has been predicted by
Goodman \& Pindor (\cite{goodman00}). Their instability works without 
self-gravity, but needs the assumption that the drag force scales non-linearily
(or more exact super-linear) with the surface density of the dust. 
Qualitatively, the mechanism works as follows (see also Goodman \& 
Pindor \cite{goodman00}): due to the friction there is a steady, but 
slow inflow of dust. If the frictional force scales super-linear with
$\Sigma_d$, the friction becomes stronger in areas with enhanced $\Sigma_d$
by this increasing the coupling to the gas and reducing the radial inflow.
As a result $\Sigma_d$ starts to grow exponentially.

   As Goodman \& Pindor (\cite{goodman00}) pointed out, the
drag description of Noh \etal (\cite{noh91}) which we used here
and which scales the frictional force strictly linear with the surface 
density does not allow for their instability mechanism. 
On the other hand, the self-gravity 
of the pressureless dust might play here a crucial role. The potential
well of a ring-like perturbation acts like a super-linear frictional term 
which reduces the inflow velocity of the dust due to its gravity: dust flows 
from outer regions to the perturbed region, but cannot leave it inwards due to 
self-gravity. Additionally, the altered local gravity decreases the 
density gradient of the gas component. Therefore, the pressure gradient
is reduced and the difference of the rotation speeds of gas and dust 
becomes smaller. As a result the frictional force and the loss of 
angular momentum decreases, by this stabilizing the formed ring.
That the dust's self-gravity is the driving agent is also supported
by the fact that the rings form, when the spatially resolved Fourier 
amplitudes become non-linear.



\subsection{Variation of dust treatment}
\label{variationdusttreatment}

\begin{figure}
   \resizebox{\hsize}{!}{
     \includegraphics[angle=270]{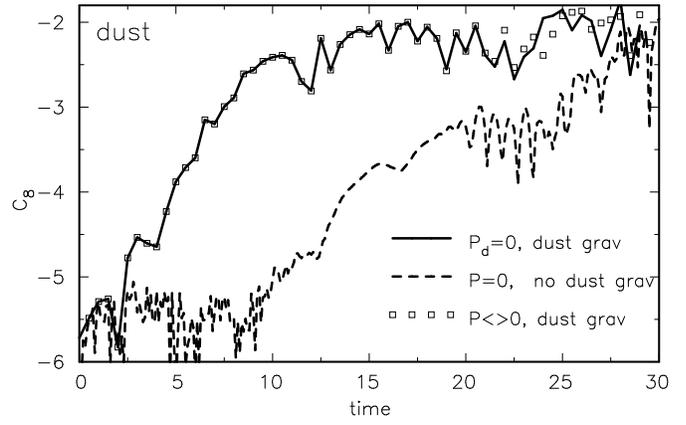}
   }
   \caption{Temporal evolution of the logarithmic Fourier amplitudes for 
    the $m=8$-mode of the dust component for different treatments
    of the dust component: reference (pressureless self-gravitating dust; 
    solid), no dust self-gravity (dashed) and non-vanishing pressure (boxes). 
    The time unit is $\natd{1.5}{7}$ yrs.}
   \label{c8treatmentdust}
\end{figure}

\begin{figure}
   \resizebox{\hsize}{!}{
     \includegraphics[angle=270]{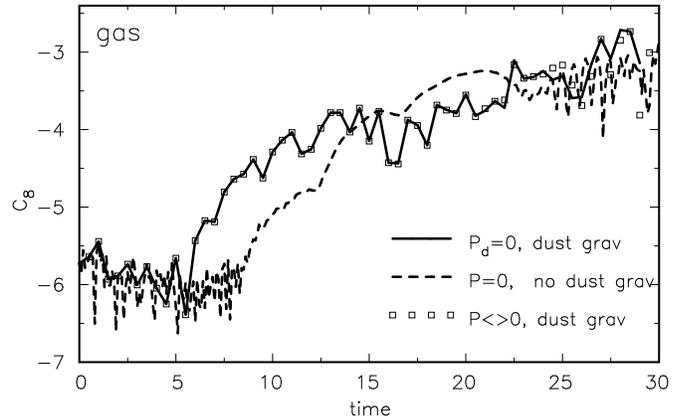}
   }
   \caption{Temporal evolution of the logarithmic Fourier amplitudes for 
    the $m=8$-mode of the gas component for different treatments
    of the dust component: reference (pressureless self-gravitating dust; 
    solid), no dust self-gravity (dashed) and non-vanishing pressure (boxes). 
    The time unit is $\natd{1.5}{7}$ yrs.}
   \label{c8treatmentgas}
\end{figure}

  As discussed at the beginning of this section, the implementation of a 
persistently pressureless component, subject to self-gravity but without
coupling to another dynamically warm component results in a very 
unstable disk.
In order to investigate the dependence of the coupled gas-dust system
on the dust treatment, we performed two experiments, in which we modified
the equation of motion of the dust by either describing the dust with
a small, but non-vanishing pressure term and by neglecting the
self-gravity of the dust.

 These simulations demonstrate that the self-gravity of the dust is the
main reason for the enhanced instability of dusty disks. In case of 
a disabled self-gravity of the dust, the dust ''feels'' only the
more slowly developing potential wells of the gaseous phase.
The Fourier amplitudes start to grow with a substantial time delay,
whereas they grow immediately in the reference model (Fig.\ 
\ref{c8treatmentdust}). Once small perturbations in the gas phase begin
to form, the pressureless dust follows them immediately as the similar 
evolution of amplitudes shows (Figs.\ \ref{c8treatmentdust} and 
\ref{c8treatmentgas}). Due to the small mass fraction of the dust, the 
growth of the gaseous density perturbations is almost unaffected by the
presence of a gravitation-free dust. In fact, the dust marginally delays 
the onset of growth of the gaseous perturbations due to the frictional force.

Different to the self-gravity of the dust component, the treatment
as a pressureless phase is less important, provided the dust is still 
dynamically cold. The growth rates of the Fourier amplitudes are merely 
identical, if an adiabatic equation of state is used for the dust as
a model with $Q_{\rm min} = 0.03$ makes clear (Fig.\ \ref{c8treatmentgas}).
The saturation levels do not vary with the different treatments of the dust.


\subsection{Dynamics of dust grains}

   For simplicity we investigated in this paper a dust component which 
is only coupled to the gas by a frictional force (and gravity). The 
collisional cross-section which enters the estimate of the frictional
timescale in the thick disk limit (cf.\ Sect.\ \ref{collisionaltimescale} and
Eq.\ (\ref{eqacollision})) is derived from the geometrical cross-section.

 Generally, it is assumed that dust particles are charged, and the sign
of the charge depends among other parameters on the grain size.
Such a charge leads to an additional Coulomb drag term for the interaction 
with ions. In case of fractional ionization ($<10^{-2}$) this contribution 
is small compared to the collisional term,
whereas it can become a substantial contribution in highly ionized areas
(Draine \cite{draine03}). In the thick-disk limit for calculating the
frictional timescale $A^{-1}$ an additional Coulomb term yields a reduction
of the frictional timescale, whereas in the thin-disk limit $A^{-1}$ 
is unaffected, because the frictional timescale is anyway dominated by 
the longer dynamical timescale.

  In addition to an enhanced Coulomb drag charged particles are also
subject to a Lorentz force exerted by the interaction with the
galactic magnetic field. 
For typical field strengths in the ISM of a few $\mu G$
Draine (\cite{draine03}) estimated a cyclotron period $\tau_{\rm B}$ 
of the order of \mbox{$\tau_{\rm B} \sim \natd{5}{4}$ yrs}.
From a comparison of $\tau_{\rm B}$ with the collisional timescale $\tau_c$ of 
a few $10^3$ yrs (cf.\ Eq.\ (\ref{eqacollisionnumbers})), we find
that the momentum exchange due to friction exceeds the coupling to the 
magnetic field, provided we apply average ISM properties for estimating 
$\tau_c$. On the other hand, in cool molecular clouds, the collisional
timescale can be longer due to the reduced thermal velocities. Then
the dust might be strongly coupled to the magnetic field.

  In addition to drag processes and a Lorentz force, the dust is also
influenced by radiation pressure due to the (usually anisotropic) interstellar
radiation field. Additionally, dust dynamics is subject to recoil
effects from photoelectric emission and photodesorption which can be
of similar order as the radiation pressure. Draine (\cite{draine03})
estimated the drift velocities related to these radiation induced effects
to be of the order of $v_{\rm drift} \sim 0.04 \, {\rm km s}^{-1}$ for a 
carbonaceous grain. Though this velocity is not very small compared to 
the relative velocities between gas and dust in our reference model 
($\sim 0.1-0.2 \, {\rm km s}^{-1}$), we neglected radiation processes, 
because the related speeds are always small compared to the velocities 
reached during the growth of instabilities. Additionally, the determination 
of the interstellar radiation field would also require a consistent treatment
of star formation and radiation transport in the ISM which is beyond the scope
of this paper. The positional error resulting from the neglection of the
drift velocities is only 4 pc within 100 Myr. 


\subsection{Miscellaneous aspects}

%
%
\subsubsection{Treatment of stellar component and dark matter}

   In our simulations stars and dark matter were treated as a rigid
component taken into account by its contribution to the rotation curve.
By this, an exchange of angular momentum between the dusty gaseous
disks and the stellar component or the halo is neglected. Such a coupling
can be very important, especially for the large-scale evolution of 
galactic disks (e.g.\ Klypin et al.\ \cite{klypin02}). 
On the other hand we focussed here mainly on the central kpc. In this 
region the mass of the dark matter is assumed to be small compared to the 
mass of the baryons (in case of normal galaxies). Thus, neglecting
an interaction between the gaseous disk and a dynamically hot halo in the
central area might be a less severe restriction. The situation is less clear 
for the stellar component. E.g.\ the perturbation exerted by a stellar bar 
will probably induce a strong $m=2$-mode which might either dominate the 
appearance of the dusty disk or lead to a complicated mode-mode coupling 
between the dominant high-$m$-modes of the dusty/gaseous disk and 
the $m=2$-bar mode.

%
%
\subsubsection{Formation and destruction of dust}

  In this work we did not consider dust formation and destruction processes. 
One reason was that we wanted to focus on the dynamical impact of the dust 
component on the stability of the gaseous disk and, thus, we wanted to 
isolate the influence as direct as possible. A second reason was that
there is still no completely satisfying model for dust formation, though
many aspects have been uncovered so far. From stellar outflows
it is known that dust formation is strongly related to
star formation; other observations suggest that dust is formed in evolved
stars or in supernovae events (e.g.\ Draine \cite{draine03}). Similarly, 
the growth and destruction of dust shows a large variety of processes 
which are important for the evolution of the dust population. Therefore, 
the implementation of a detailed evolutionary model for the dust component
seems to be out of reach at the moment. On the other hand, a simplified
treatment of dust formation and destruction might be possible on the basis
of a live stellar component which includes a simple description of star 
formation and stellar evolution. 

   Due to the complexity of the processes governing the formation and
destruction of dust, only rough estimates on the related timescales
are possible. If we assume that dust formation is mainly driven by stars
(or supernovae), then the star formation timescale $\tau_{\rm SF}$ 
provides a lower limit for the dust formation timescale $\tau_{\rm df}$.
For normal spiral galaxies the star formation rate is typically of the 
order of one or a few $\msun/yr$ yielding a timescale of the order of 
(several) $10^9$ yrs. A similar lifetime was invoked by Draine \& Salpeter 
(\cite{draine79}) in order to explain the large fraction of Si bound in 
dust grains (and not in the gaseous ISM). These lifetimes exceed
other involved timescales like the revolution periods or the frictional 
timescales by far. Thus, dust formation and destruction is probably 
less important for the overall dynamical processes presented in this paper.
On the hand, the secular evolution of the disk might be strongly affected, 
when the dust-to-gas ratio becomes sufficiently large. 
Additionally, the spatial distribution of the dust-to-gas ratio might
be influenced by locally concentrated star forming events. 

These mentioned aspects (a live stellar disk, the influence of a stellar 
(mini-)bar and effects related to simplified dust formation and destruction
model) will be part of a future paper.


\section{Summary}
\label{summary}

   We investigated the influence of a cold dust component on the evolution
of galactic gaseous disks by means of 2-dimensional hydrodynamical simulations
for flat disks. We focussed especially on the question, whether a small 
contribution of dust is able to destabilize a gaseous disk and what kind 
of structures are formed. 
We coupled gas and dust by a drag force depending on the relative 
velocities between both components and a frictional timescale. The latter 
is derived either only from collisions between gas and dust
particles (thick disk limit) or from collisions followed by 
momentum equipartition in the gaseous disk (thin disk limit). 
Our initial models are composed of a submaximal exponential
gaseous disk, a small admixture of dust and a rigid stellar and dark halo
component. The rotation curve is assumed to rise linearily inside 
a radius of 100 pc becoming flat outside.

   From the evolution of the Fourier amplitudes we found that the 
higher-order modes are the dominant unstable modes in the purely
gaseous simulation (in our case $m=8$). Their growth is mainly restricted 
to the central kpc. An admixture of 2\% dust (relative to the gas mass) 
destabilizes the gaseous disk. The dust component becomes non-linear 
after 100 Myr, followed by the gas 50 Myrs later. Different to the dust-free
calculations, all modes have similar Fourier amplitudes.

  The formed structures of both, gas and dust, are rather patchy and
multi-armed as expected from the similar Fourier amplitudes of the 
different modes. The dust component shows much more contrast between
arm and inter-arm regions than the gas. The dust is spatially correlated 
with gas, but it does not exactly follow the gas. Peaks in the dust 
distribution are frequently at the inner edges of peaks in the gas 
distribution. Some dust peaks are also completely outside gas concentrations
and some are exactly at the positions of gas peaks. This results in a large
scatter of dust-to-gas ratios at different places. The dust develops also
thin filaments which sometimes connect the arms. Therefore, the dust
distribution has a more cellular appearance, whereas the gas develops
a multi-armed spiral morphology. The dust is sometimes organized in a 
ring-like structure which is the result of an instability driven by
the frictional force and the self-gravity of the dust (or adjacent gas).

  Below a dust-to-gas mass ratio $r$ of 1\% the dynamical influence of 
the dust on the gaseous disk becomes negligible. This critical value is 
close to the observed mean value in normal galaxies like the Milky Way. 
Since the dust-to-gas ratio scales linearily with metallicity, larger local
values of $r$, especially in the central galactic regions, seem to be
reasonable. Such values are also in agreement with local gas-to-dust
determinations. Since already a dust-to-gas ratio of 2\% significantly 
affects the evolution of the disk, even the observed small dust admixtures
are expected to have an impact on the dynamics of some galaxies 
(e.g.\ the dust-rich M51). For a 10\% admixture of dust the gaseous 
component is completely destabilized. The growth rates are enhanced by a 
factor of 3-4 with almost no latency phase. The saturation levels
reached after 30 Myr are substantially larger than in the low-$r$ models. 

  The Toomre parameter $Q$ of the gaseous disk has almost no influence on the 
dust component, but it strongly controls the gaseous phase. As usual 
a larger $Q$ gives more stable disks. However, the destabilizing influence
of the cold dust was even found for a hot disk with $Q=3$, though the
saturation level is too small to be observable. 

The results are robust with respect to technical parameters like application
of artificial viscosity, grid size and grid boundaries. They are also
independent on the adopted equation of state for the gas. For the dust
treatment it is essential to take the self-gravity of the dust into account.
The adopted equation of state of the dust is less important, provided the
gas is treated as a dynamically cold component. 


\begin{acknowledgements}

   N.O. acknowledges financial support from the Deutsche 
Forschungsgemeinschaft under grant 436 RUS 17/65/02 which made a visit in Kiel
possible. The simulations have been performed on the NEC-SX5 at the 
computing center of the university of Kiel. The authors are
grateful to Vladimir Korchagin for discussions about disk stability
which stimulated this work.

\end{acknowledgements}


\end{document}